\documentclass[12pt]{article}
\usepackage{graphicx} 
\RequirePackage{amsthm,amsmath,amsfonts,amssymb}
\usepackage{natbib}
\usepackage{url} 
\usepackage{algorithm}
\usepackage{algpseudocode}
\usepackage{setspace}
\usepackage[
  colorlinks=true,
  linkcolor=blue,
  citecolor=blue,
  urlcolor=blue
]{hyperref}
\usepackage{footmisc} 
\usepackage{float}
\usepackage{fullpage}
\usepackage{tikz}
\usetikzlibrary{positioning, quotes, trees, shapes.geometric}
\usepackage{standalone}
\usepackage{comment}
\usepackage{subcaption}
\usepackage{booktabs}
\usepackage{xcolor}
\usepackage{multirow}
\usepackage{booktabs}
\usepackage{makecell}
\usepackage{csquotes}
\usepackage{titlesec}
\usepackage{amsthm}

\DeclareRobustCommand{\ellipsenode}{%
  \tikz[baseline=-0.6ex] 
    \draw[line width=0.6pt, fill=white] (0,0) ellipse (0.4em and 0.2em);%
}

\DeclareRobustCommand{\rectnode}{%
  \tikz[baseline=-0.1ex] 
    \draw[line width=0.6pt, fill=white] (0,0) rectangle (.8em, .5em);%
}

\DeclareRobustCommand{\roundrectnode}{%
  \tikz[baseline=-0.1ex]
    \draw[line width=0.6pt, rounded corners=2pt, fill=white]
      (0,0) rectangle (.8em,0.5em);%
}

\usepackage{xr-hyper}
\usepackage{hyperref}
\begin{document}


\title{\textbf{False Positives, False Negatives, and the Detection-Only Problem: A Hierarchical Model for Species Occurrence with Observation Error}}

\author{
  {\normalsize Kabiru Abubakari$^{1,3}$},\;
  {\normalsize Eleni Matechou$^{1}$},\;
  {\normalsize Marie C. Henniges$^{2,3}$},\,
  {\normalsize Ilia J. Leitch$^{3}$},\\
  {\normalsize Andrew R. Leitch$^{2}$} \&\;
  {\normalsize Silvia Liverani$^{1}$}
}

\footnotetext{%
  \noindent\leavevmode\\[-\baselineskip]
  $^{1}$School of Mathematical Sciences, Queen Mary University of London, United Kingdom.\\
  $^{2}$School of Biological and Behavioural Sciences, Queen Mary University of London, United Kingdom.\\
  $^{3}$Royal Botanic Gardens, Kew, Richmond, Surrey, United Kingdom.
}
\date{}

\maketitle
\begin{abstract}

Monitoring species occurrence is essential for understanding biodiversity change, informing 
conservation decisions, and assessing the impact of environmental pressures on ecosystems. 
Species occurrence data arise from different survey designs, and the statistical literature 
has developed distinct corresponding modelling approaches --- namely occupancy models, species distribution 
models, and presence-only methods --- whose fundamental connections have remained largely 
unrecognised. We argue that these are all special cases of a single hierarchical observation process. 
To make these connections explicit, we introduce a unified terminology centred on two 
data types: detection/non-detection data with $T$ visits (\text{DN-}$T$) and 
detection-only data (\text{DO}) --- where \text{DN-}$T$ with $T>1$ corresponds to 
traditional occupancy modelling, \text{DN-1} to species distribution modelling, and 
\text{DO} to what the literature commonly, but we argue inaccurately, calls 
presence-only data. Within 
this framework, we study the identifiability of \text{DO} models and propose a novel 
hierarchical model for \text{DO} data that, for the first time, explicitly 
accounts for both false positive and false negative detection errors. Identifiability is 
achieved through prior distributions that express the natural belief that a species is more likely to be 
recorded where it is present than where it is absent. Simulation studies demonstrate that 
existing approaches, which ignore observation error, yield severely biased estimates of the 
coefficients of environmental conditions on occurrence probability, with credible interval 
coverage collapsing to zero, while our model achieves estimation performance comparable 
to \text{DN-1} --- despite \text{DO} data being substantially less structured and less 
informative. We further show that the well-known divergence of the intercept as background sample size grows — a recognised pathology of existing approaches — is a direct consequence of ignoring observation error, and disappears naturally under our model. The framework is 
illustrated using vascular plant data from New Zealand, where \text{DO} and \text{DN-1} 
results are directly compared, and from the United Kingdom, where only \text{DO} data 
are available.
\end{abstract}

\section{Introduction}\label{sec:intro}

Biodiversity is declining at an unprecedented rate, driven by habitat loss, climate 
change, invasive species, and other anthropogenic pressures 
\citep{ceballos2017, ipbes2019, urban2015}. Understanding how species respond to these 
pressures, and anticipating future changes in their distributions, is fundamental to 
designing effective conservation strategies, prioritising protection efforts, and assessing 
the ecological consequences of environmental change \citep{thuiller2008predicting, 
thuiller2009biomod}. Central to this endeavour is the ability to model species occurrence 
across space --- to identify which environmental conditions drive a species to be present 
or absent, and to predict occurrence across unsampled locations. It is therefore crucial 
that the statistical methods used to analyse species occurrence data yield reliable 
estimates of the effect of environmental conditions on occurrence probability.

Species occurrence data arise from different survey designs, and the statistical 
literature has developed distinct modelling approaches in response, whose fundamental 
connections have remained largely unrecognised. When a study area is well-defined and 
sites are visited repeatedly, both detections and non-detections are 
recorded; models developed for such data are known as occupancy models 
\citep{mackenzie2002estimating, royle2006generalized, mackenzie2017occupancy}. When sites 
are visited only once within a structured survey, the corresponding models are typically referred to in the 
literature as species distribution models \citep{lele2012dealing, solymos2012conditional}. 
Finally, when data are collected opportunistically --- through citizen science activities, museum 
records, herbarium collections, and online biodiversity portals such as the Global 
Biodiversity Information Facility (GBIF) --- only species detections are recorded, and 
such data have been modelled using what the literature calls presence-only methods 
\citep{phillips2004maximum, phillips2008modeling, warton2010poisson, elith2006novel}. 
Despite being treated as fundamentally different problems requiring distinct methodological 
traditions, we argue that occupancy models, species distribution models, and presence-only 
methods are all special cases of a single hierarchical observation process. To make these 
connections explicit, we introduce a unified terminology: detection/non-detection data with 
$T$ visits (\text{DN-}$T$) and detection-only data (\text{DO}). Under this terminology, 
\text{DN-}$T$ with $T > 1$ corresponds to occupancy modelling, \text{DN-1} to 
species distribution modelling, and \text{DO} to what the literature calls presence-only 
modelling --- a misnomer, since, as we argue in the paper, a recorded detection is not necessarily a confirmed 
presence.

Within this unified framework, we focus in particular on \text{DO} data, which are 
increasingly abundant yet present the greatest statistical challenges. The integration of 
species recognition software in mobile phone applications such as iNaturalist and PlantNet 
has dramatically boosted citizen science activity, increasing the volume of detection 
records available through portals such as GBIF and the Botanical Society of Britain and 
Ireland (BSBI). \text{DO} data arise in two distinct settings with different implications 
for modelling: those collected across a well-defined area with a predetermined number of 
sites, such as the BSBI grid of the United Kingdom where each grid square carries a unique 
identifier, and those arising from entirely incidental observation where the set of surveyed 
sites is not defined in advance, such as records held by GBIF. In both settings, a missing record at a site is fundamentally ambiguous: it may reflect no visit having taken place, a visit at which the species was absent and correctly went unrecorded, or a visit at which the species was present but was not detected. Equally, a detection record is not unambiguous evidence of presence: it may reflect a true detection or a false positive error whereby the species was recorded at a site it does not occupy. This ambiguity has serious consequences 
for inference on the drivers of species distributions \citep{syfert2013}.

Various modelling approaches have been proposed for \text{DO} data, including the maximum 
entropy model \citep[Maxent;][]{phillips2004maximum, phillips2006maximum, 
phillips2008modeling}, the inhomogeneous Poisson point process model 
\citep[IPPP;][]{warton2010poisson, chakraborty2011point, moreira2022analysis}, logistic 
regression with background samples \citep{ward2009presence, dorazio2012predicting}, and a 
wide range of machine learning approaches including boosted regression trees, random 
forest, and extreme gradient boosting \citep{elith2006novel, valavi2022predictive}. 
Despite their methodological differences, these approaches share a fundamental limitation: 
none accounts for observation errors --- that is, for the possibility of both false 
positive and false negative detections --- and this leads to biased estimates of the 
coefficients of environmental conditions on occurrence probability, as we demonstrate in Section~\ref{sec:simstudy}. Attempts to model the detection process within the IPPP framework require separate, independent covariates for the occurrence and detection processes for model identification \citep{dorazio2012predicting, fithian2012finite}; and whilst \cite{moreira2022analysis} demonstrate that parameter identifiability is achievable in this setting, they acknowledge that covariates for the detection process are not readily available for \text{DO} data in practice, limiting the applicability of this approach. At the same time, structured repeated surveys that would yield the 
richer \text{DN-}$T$ data required by occupancy models are costly \citep{bowler2026predicting} 
and can disturb the species being studied \citep{lele2012dealing}, making the development 
of principled methods for the widely available \text{DO} data all the more pressing. A further limitation shared by virtually all existing approaches is the assumption 
of perfect or near-perfect detection: standard occupancy models typically assume no 
false positive errors \citep{mackenzie2002estimating}, species distribution models 
and machine learning approaches treat observed records as direct realisations of the 
latent occurrence state, and \text{DO} methods assume that detections correspond to 
true presences. Whilst \cite{royle2006generalized} relaxed this assumption for 
\text{DN-}$T$ data by allowing both false positive and false negative errors 
simultaneously, no existing method does so for \text{DO} data.

We propose a novel hierarchical model for \text{DO} data that, for the 
first time, explicitly accounts for both false positive and false negative detection errors. 
The key insight is that the species observation process underlying \text{DO} data is 
structurally analogous to that of \text{DN-1}, but with the probability of site visit 
unknown and confounded with the detection probabilities, resulting in substantially greater 
uncertainty. We study the identifiability of \text{DO} models within this framework and 
show that it is achieved through prior distributions that reflect the natural and ecologically justifiable belief that the probability of recording a species at a site it occupies is greater than the probability of recording it at a site it does not occupy. This avoids the need for separate detection covariates, 
which are not available for \text{DO} data.

The benefits of this approach are demonstrated through extensive simulation studies. 
Existing approaches that ignore observation error yield severely biased estimates of the 
coefficients of environmental conditions on occurrence probability, with credible interval 
coverage collapsing to zero across all simulation scenarios considered. Our model achieves 
estimation performance comparable to \text{DN-1}, despite \text{DO} data being 
substantially less structured and less informative: \text{DN-1} data record both detections 
and non-detections from planned visits, whereas \text{DO} data record only detections from 
visits of unknown probability. We further show that accounting for observation error 
resolves the well-known divergence of the intercept term as background sample size grows 
\citep{owen2007infinitely, renner2013equivalence}, a phenomenon that arises under naive 
model specification and disappears under correct specification within our framework. The 
approach is illustrated using vascular plant data from New Zealand, where \text{DO} and 
\text{DN-1} results are directly compared using a well-established benchmark dataset 
\citep{elith2020presence}, and from the United Kingdom using \text{BSBI} records 
\citep{henniges2022taxonomic}, where only \text{DO} data are available.

The rest of the paper is organised as follows. Section~\ref{bg} presents the background 
and review of existing methods, establishing the connections between \text{DO} and 
\text{DN-}$J$ models within the unified framework. Section~\ref{sec:modelling} presents 
the proposed hierarchical model for \text{DO} data, including the likelihood, prior 
distributions, and identifiability analysis. Section~\ref{sec:simstudy} presents the 
simulation study. Section~\ref{sec:casestudy} presents the two case studies, and 
Section~\ref{sec:discussion} concludes with a discussion of findings, limitations, and 
directions for future work.
\section{Background}\label{bg}

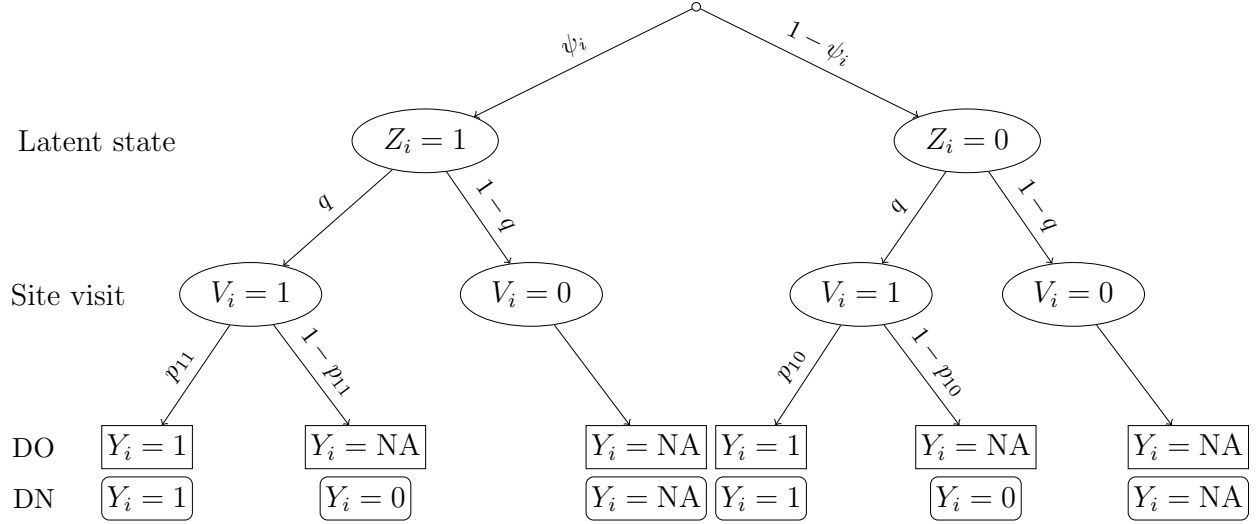
\begin{figure}[!ht]
  \centering
  \begin{tikzpicture}[
    node distance = 10mm and 20mm, 
    dot/.style={circle, fill=white, inner sep=1.2, minimum size=1mm},
    H/.style = {ellipse, draw, fill=white, minimum width=15mm, minimum height=8mm}, 
    Y1/.style={rectangle, draw, fill=white, minimum width=12mm, minimum height=6mm, inner sep=2pt},
    Y2/.style={rectangle, draw, fill=white, minimum width=12mm, minimum height=6mm, inner sep=2pt, rounded corners=3pt},
    every edge quotes/.style = {auto, font=\footnotesize, sloped},
    row label/.style = {anchor=east, font=\small, xshift=-5mm}]

    \begin{scope}[nodes = H]
        \node[dot] (1) {};
        \node[H] (2) [below left=15mm and 30mm of 1] {$Z_i = 1$};
        \node[H] (3) [below right=15mm and 30mm of 1] {$Z_i = 0$};
        \node[H] (4) [below left=15mm and 10mm of 2] {$V_i = 1$};
        \node[H] (5) [below right=15mm and 0.5mm of 2] {$V_i = 0$};
        \node[Y1] (6) [below left=15mm and 1mm of 4] {$Y_i = 1$};
        \node[Y1] (7) [below right=15mm and 0.5mm of 4] {$Y_i = \text{NA}$};
        \node[Y1] (8) [below right=15mm and 0.5mm of 5] {$Y_i = \text{NA}$};
        \node[H] (9) [below left=15mm and 0.5mm of 3] {$V_i = 1$};
        \node[Y1] (10) [below left=15mm and 0.5mm of 9] {$Y_i = 1$};
        \node[Y1] (11) [below right=15mm and 0.5mm of 9] {$Y_i = \text{NA}$};
        \node[H] (12) [below right=15mm and 0.5mm of 3] {$V_i = 0$};
        \node[Y1] (13) [below right=15mm and 0.5mm of 12] {$Y_i = \text{NA}$};
        \node[Y2] (14) [below=1mm of 6] {$Y_i = 1$};
        \node[Y2] (15) [below=1mm of 7] {$Y_i = 0$};
        \node[Y2] (16) [below=1mm of 8] {$Y_i = \text{NA}$};
        \node[Y2] (17) [below=1mm of 10] {$Y_i = 1$};
        \node[Y2] (18) [below=1mm of 11] {$Y_i = 0$};
        \node[Y2] (19) [below=1mm of 13] {$Y_i = \text{NA}$};
        
    \end{scope}

    \draw[->] (1) edge["$\psi_i$"] (2);
    \draw[->] (1) edge["$1 - \psi_i$"] (3);
    \draw[->] (2) edge["$q$"] (4);
    \draw[->] (2) edge["$1 - q$"] (5);
    \draw[->] (4) edge["$p_{11}$"] (6);
    \draw[->] (4) edge["$1 - p_{11}$"] (7);
    \draw[->] (5) edge[] (8);
    \draw[->] (3) edge["$q$"] (9);
    \draw[->] (9) edge["$p_{10}$"] (10);
    \draw[->] (9) edge["$1-p_{10}$"] (11);
    \draw[->] (3) edge["$1 - q$"] (12);
    \draw[->] (12) edge[] (13);

    \node[row label] at ([yshift=0mm]6.west) {DO};
    \node[row label] at ([yshift=0mm]14.west) {DN};
    \node[anchor=east, xshift=-23mm] at (2.west) {Latent state};
    \node[anchor=east, xshift=-6mm] at (4.west) {Site visit};

\end{tikzpicture}
\caption{\textit{A hierarchical illustration of the species observation process 
underlying all occurrence data, regardless of survey design. 
\textup{(\ellipsenode)} represents latent states that are never directly 
observed: the occupancy state $Z_i \in \{0, 1\}$, indicating whether the 
species occupies site $s_i$, and the site visit indicator $V_i \in \{0, 1\}$, 
indicating whether the site was visited, assumed independent of $Z_i$. 
\textup{(\rectnode)} represents detection-only data \textup{(DO)}, arising 
from unstructured surveys in which $V_i$ is unknown and only detections 
($Y_i = 1$) are recorded; a missing record ($Y_i = \text{NA}$) is ambiguous, 
as it may reflect no visit having taken place, a visit at which the species 
was absent and correctly went unrecorded, or a visit at which the species was 
present but was not detected. \textup{(\roundrectnode)} represents 
detection/non-detection data for a single visit \textup{(DN-1)}, arising from 
structured surveys in which $V_i = 1$ for all sites in the sample by design 
and both detections ($Y_i = 1$) and non-detections ($Y_i = 0$) are recorded. 
Detection/non-detection data with $T > 1$ visits \textup{(DN-}$T$\textup{)} 
arise by repeating the \textup{DN-1} process independently across $T$ visits. 
In all survey designs, a detection may arise either because the species truly 
occupies the site and is correctly recorded (with probability $p_{11}$), or 
because of a false positive error whereby the species is recorded at a site it 
does not occupy (with probability $p_{10}$). The key distinction between survey 
designs is therefore what is observed: \textup{DO} surveys record only 
detections, \textup{DN-1} surveys record both detections and non-detections 
from a single planned visit, and \textup{DN-}$T$ surveys record both detections 
and non-detections across $T$ planned visits. All three arise from the same 
underlying hierarchical process, differing only in what is observed and what 
remains latent.}}
  \label{fig:tree}
\end{figure}

This section formalises the unified hierarchical observation process introduced in 
Section~\ref{sec:intro}. We begin by defining the notation and the data-generating 
process that underlies all species occurrence data, regardless of survey design, and 
show how Figure~\ref{fig:tree} summarises this process. We then review existing 
statistical methods for \text{DN-}$T$ data in Section~\ref{bg1.1} and for \text{DO} 
data in Section~\ref{bg1.3}, identifying in each case what is and is not accounted 
for in terms of observation error. Section~\ref{bg1.4} establishes the formal 
connections between these methods, motivating the novel model proposed in 
Section~\ref{sec:modelling}.

Let $\mathcal{A} \subseteq \mathbb{R}^2$ be the spatial domain of interest, with 
sites $\left\{s_i : i = 1, 2, \dotsc, n \right\}$. At each site $s_i$, the latent 
occupancy state $Z_i \in \left\{0, 1\right\}$ indicates whether the species occupies 
site $s_i$; this state is never directly observed. A vector of environmental 
conditions $\mathbf{x}_i$ is measured at each site. The probability of species presence 
$\psi_i = \mathbb{P}(Z_i = 1 \mid \mathbf{x}_i)$ is modelled as a function of 
$\mathbf{x}_i$ via the logistic link
\begin{equation}
    \psi_i = \text{logit}^{-1}\left\{ \eta_i(\mathbf{x})\right\}; \quad 
    \eta_i(\mathbf{x}) = \mathbf{x}_i^\top\boldsymbol{\beta},
    \label{eq:psi}
\end{equation}
where $\boldsymbol{\beta}$ contains the intercept and the coefficients of the 
environmental conditions, which are the primary objects of inference throughout 
this paper. Given $\psi_i$, the latent occupancy state is distributed as
\begin{equation}
    Z_{i} \sim \text{Bernoulli}(\psi_i).
    \label{eq:latent}
\end{equation}

For \text{DO} data, it is important to distinguish between two settings that have 
different implications for how the likelihood is constructed. In the first setting, 
the study area $\mathcal{A}$ is discretised into a predetermined grid of $n$ sites, 
each with a unique identifier, so that $n$ is fixed and known. For a given period, 
each site either has a detection record ($Y_i = 1$) or no record ($Y_i = \text{NA}$), 
and the likelihood can be constructed directly over all $n$ sites. The UK case study 
in Section~\ref{sec:casestudy} is of this type, where the BSBI has discretised 
the map of the United Kingdom into hectads. In the second setting, the set of sites 
is not defined in advance and $n$ is not known: records arise from entirely 
incidental observation, and we only know the locations of detections. In this case, 
background samples drawn from across $\mathcal{A}$ are required to represent the 
unsampled part of the study area and to approximate the integral over $\mathcal{A}$ 
in the likelihood. The New Zealand case study in Section~\ref{sec:casestudy} is 
of this type. The data-generating process illustrated in Figure~\ref{fig:tree} 
applies to any individual site $s_i$ and is identical in both settings; the 
distinction arises only in how the likelihood is constructed across sites, as we 
discuss in Section~\ref{sec:modelling}.

What is observed at site $s_i$ depends on the survey design, and it is this that 
distinguishes the three data types within our unified framework. In all cases, 
whether a site is visited is governed by a visit process: we denote by $V_i \in 
\{0, 1\}$ the indicator that site $s_i$ was visited, and assume that visits occur 
independently of the latent occupancy state $Z_i$. Observation error arises 
conditional on a visit having taken place: a species may be falsely recorded as 
present at an unoccupied site (false positive error, with probability $p_{10}$), 
or fail to be recorded at an occupied site (false negative error, with probability 
$p_{01} = 1 - p_{11}$, where $p_{11}$ is the probability of a correct detection 
at an occupied site). The key distinction between survey designs is whether $V_i$ 
is known and whether non-detections are recorded. Figure~\ref{fig:tree} illustrates 
this data-generating process in full, showing how \text{DN-}$T$ and \text{DO} data 
both arise as partial observations of the same underlying latent process: in 
\text{DN-}$T$ surveys, $V_i = 1$ for at least one visit per site in the sample by design and both 
detections ($Y_i = 1$) and non-detections ($Y_i = 0$) are recorded; in \text{DO} 
surveys, $V_i$ is unknown for sites with no detections, and only detections are recorded, so that a missing record 
($Y_i = \text{NA}$) is fundamentally ambiguous --- it may reflect no visit having 
taken place, a visit at which the species was absent and correctly went unrecorded, 
or a visit at which the species was present but was not detected. The remainder of 
this section reviews existing methods for each data type in turn.
\titleformat{\subsection}
  {\normalfont}   
  {\thesubsection}        
  {0.5em}                   
  {\itshape}  

\titlespacing*{\subsection}
  {2em}                  
  {1.5ex plus 1ex minus .2ex}  
  {1.5ex plus .2ex}      

\titleformat{\subsubsection}
  {\normalfont}           
  {\thesubsubsection}     
  {0.5em}                 
  {\itshape}              

\titlespacing*{\subsubsection}
  {3em}                   
  {1ex plus 0.5ex minus 0.2ex}  
  {1ex plus 0.2ex}        
  
\subsection{Detection/non-detection data with $T > 1$ visits 
            \textup{(DN-}$T$\textup{)}}\label{bg1.1}

For \text{DN-}$T$ data, the visit indicator $V_{it} \in \{0, 1\}$ is known for 
all sites $i$ and occasions $t$, and each site must be visited on at least one 
occasion, independently of $Z_i$. When $V_{it} = 1$, the observed data 
$Y_{it} \in \{0, 1\}$ arise from the observation model \citep{mackenzie2002estimating, 
royle2006generalized}
\begin{equation}
    \begin{aligned}
     Y_{it} \mid Z_i = 1, V_{it} = 1 &\sim \text{Bernoulli}(p_{11t}), \\
     Y_{it} \mid Z_i = 0, V_{it} = 1 &\sim \text{Bernoulli}(p_{10t}),
    \end{aligned}
    \label{eq:obs_dnt}
\end{equation}
where $p_{11t}$ and $p_{10t}$ are the true positive and false positive detection 
probabilities at occasion $t$ respectively, allowed to, but not required to, vary across occasions for example as a function of covariates within a logistic regression model. When 
$V_{it} = 0$, $Y_{it}$ is not observed. Given $Z_i$, observations $Y_{it}$ and 
$Y_{it'}$ are conditionally independent for $t \neq t'$, as each arises 
independently from the observation model in Equation~\ref{eq:obs_dnt}. The latent occupancy state $Z_i$ is assumed fixed across all occasions --- the 
closed-population assumption \citep{mackenzie2017occupancy} --- which is necessary 
for the model to be well-defined across multiple visits and for $\psi_i$ to have 
a consistent interpretation. The log-likelihood for \text{DN-}$T$ and full 
technical details are provided in Supplementary Material~\ref{supp:dnt}.

The known visit structure in \text{DN-}$T$ data, combined with observations across 
multiple occasions, provides sufficient information to estimate $p_{11t}$ and 
$p_{10t}$ alongside $\boldsymbol{\beta}$, without requiring a separate covariate 
set for the detection process. However, the likelihood exhibits a symmetry: swapping 
the roles of $p_{11t}$ and $p_{10t}$ and replacing $\psi_i$ with $1 - \psi_i$ 
yields an identical likelihood value \citep{royle2006generalized, de2021need}, 
leading to a bimodal posterior. This is resolved by imposing the constraint 
$p_{11t} > p_{10t}$ for all $t$ \citep{royle2006generalized, guillera2017dealing, 
griffin2020modelling}: it is more likely to detect a species at a site it occupies 
than to record it falsely at a site it does not, a condition that is both 
ecologically justifiable and ensures that the coefficients $\boldsymbol{\beta}$ 
retain their correct ecological interpretation. This constraint is adopted 
throughout the paper and is central to achieving identifiability in the proposed 
\text{DO} model in Section~\ref{sec:modelling}.

The most widely used \text{DN-}$T$ models assume $p_{10t} = 0$ for all $t$
\citep{mackenzie2002estimating, mackenzie2017occupancy} --- that is, all detections 
are true presences and only false negative errors are possible. 
\cite{royle2006generalized} relaxed this assumption to allow $p_{10t} > 0$, 
permitting both false positive and false negative errors simultaneously. As we 
discuss in Section~\ref{bg1.3}, this more general formulation is directly analogous 
to the observation model we propose for \text{DO} data, and the constraint 
$p_{11t} > p_{10t}$ plays the same identifiability role in both settings.

\subsection{Detection/non-detection data with $T=1$ visit 
            \textup{(DN-1)}}\label{bg1.2}

For \text{DN-1} data, $V_i = 1$ for all sites $i$ and the visit indicator is 
known, as in \text{DN-}$T$, but each site is visited on a single occasion only. 
The observed data $Y_i \in \{0, 1\}$ follow the same observation model as 
Equation~\ref{eq:obs_dnt} with the occasion index dropped, so that both 
detections ($Y_i = 1$) and non-detections ($Y_i = 0$) are recorded, as 
illustrated in Figure~\ref{fig:tree}. The closed-population assumption is 
trivially satisfied with a single visit. This is the data type that the 
literature typically refers to as arising from species distribution models 
\citep{lele2012dealing, solymos2012conditional}.

The log-likelihood for \text{DN-1} is
\begin{equation}
\ell(p_{11},\,p_{10},\,\boldsymbol{\beta} \mid \mathbf{Y})
= \sum_{i=1}^{n} 
\log \Big[ 
\big(p_{11}\psi_i + p_{10}(1-\psi_i)\big)^{Y_i} 
\big(1 - p_{11}\psi_i - p_{10}(1-\psi_i)\big)^{1-Y_i} 
\Big],
\label{eq:dn1lik}
\end{equation}
which is a Bernoulli mixture with parameters $p_{11}$ and $p_{10}$ for occupied 
and unoccupied sites respectively. The connection between this likelihood and 
that of the proposed \text{DO} model is established in Section~\ref{sec:modelling}.

The most widely used approaches for \text{DN-1} data --- logistic regression, 
random forest, boosted regression trees, and other machine learning methods 
\citep{elith2006novel, cutler2007random, elith2008working, valavi2022predictive} 
--- treat $Y_i$ as if corresponds to $Z_i$, implicitly assuming $p_{11} = 1$ 
and $p_{10} = 0$: every detection is a true presence and every non-detection is a 
true absence. This is the naive approach, and whilst it may yield adequate 
predictive performance in some settings, it produces biased estimates of 
$\boldsymbol{\beta}$ whenever observation error is non-negligible, as we 
demonstrate in Section~\ref{sec:simstudy}.

Principled \text{DN-1} models that account for observation error include the 
single-visit occupancy model of \cite{lele2012dealing}, which imposes $p_{10} = 0$ 
and requires a separate covariate set $\mathbf{v}_i$ for detection, and the 
conditional likelihood approach of \cite{solymos2012conditional}, which similarly 
requires $\mathbf{x}_i$ and $\mathbf{v}_i$ to be sufficiently different for 
identifiability. In both cases, suitable detection covariates are rarely available 
in practice \citep{moreira2022analysis}, limiting the applicability of these 
approaches. The constraint $p_{11} > p_{10}$ established in Section~\ref{bg1.1} 
applies equally here and, as we show in Section~\ref{sec:modelling}, informative 
prior distributions on $p_{11}$ and $p_{10}$ provide a practical route to 
identifiability that avoids the need for separate detection covariates entirely.

\subsection{Detection-only data \textup{(DO)}}\label{bg1.3}

For \text{DO} data, at sites where a detection is recorded ($Y_i = 1$), the 
visit indicator $V_i = 1$ is known, since a detection necessarily implies a 
visit. At sites with no record ($Y_i = \text{NA}$), however, $V_i$ is unknown: 
the missing record could reflect no visit having taken place, a visit at which 
the species was absent and correctly went unrecorded, or a visit at which the 
species was present but was not detected, as illustrated in 
Figure~\ref{fig:tree}. The probability of a site being visited, $q$, is unknown 
throughout --- this is the fundamental distinction from \text{DN-}$T$ and 
\text{DN-1} data, where the visit structure is known. For \text{DO} data, however, $q$ is not purely a visit probability: it absorbs all processes that determine whether a site generates a detection record, including reporting behaviour, recorder interest, and habitat accessibility. These factors are inseparable from the visit process in \text{DO} data, and their conflation is an inherent feature of the data type rather than a modelling limitation. The observation model conditional on a record being generated is the same as in 
Equation~\ref{eq:obs_dnt}: a detection arises either because the species occupies 
the site and is correctly recorded (with probability $p_{11}$), or because of a 
false positive error (with probability $p_{10}$). However, because $q$ is unknown 
and absorbs visit probability, reporting behaviour, and habitat accessibility 
simultaneously, $p_{11}$ and $p_{10}$ are confounded with $q$ and only the products 
$\theta_{11} = qp_{11}$ and $\theta_{10} = qp_{10}$ are estimable from \text{DO} 
data alone, as we discuss in detail in Section~\ref{sec:modelling}.

As discussed above, the likelihood for \text{DO} data is constructed differently 
depending on whether $n$ is fixed or unknown. When $n$ is fixed, all $n$ sites 
are known: at the $n_1$ detection sites $Y_i = 1$ and $V_i = 1$, whilst at the 
remaining $n - n_1$ sites $Y_i = \text{NA}$ and $V_i$ is unknown. The likelihood 
is a direct product over all $n$ sites, with each $Y_i = \text{NA}$ contributing 
the probability of not obtaining a detection at site $s_i$, which integrates over 
all three possibilities for a missing record. When $n$ is unknown, only the $n_1$ 
detection sites are observed and background samples drawn from across $\mathcal{A}$ 
are required to approximate the integral over the study area in the likelihood. 
In both cases the observation model is identical and the same inferential 
challenges arise, as we discuss in Section~\ref{sec:modelling}.

Various modelling approaches have been proposed for \text{DO} data. Maxent 
\citep{phillips2004maximum, phillips2006maximum, phillips2008modeling} estimates 
the distribution of environmental conditions at detected sites relative to the 
study area, and is equivalent to an inhomogeneous Poisson point process model 
\citep[IPPP;][]{warton2010poisson, renner2013equivalence}. Logistic regression 
with background samples \citep{ward2009presence, dorazio2012predicting} and 
machine learning approaches including boosted regression trees, random forest, 
and extreme gradient boosting \citep{elith2006novel, valavi2022predictive} treat 
\text{DO} data as a binary classification problem. Despite their methodological 
differences, all of these approaches share the same fundamental limitation within 
our unified framework: they assume that detections correspond to true presences, 
thereby ignoring false positive errors, and none accounts explicitly for false 
negative errors in the \text{DO} data. Whilst the IPPP can in principle model 
both the occurrence and detection process as a thinned Poisson process 
\citep{chakraborty2011point}, doing so requires separate, independent covariate 
sets for the occurrence and detection processes \citep{dorazio2012predicting, 
fithian2012finite}; and whilst \cite{moreira2022analysis} demonstrate that 
identifiability is achievable in this setting, they acknowledge that detection 
covariates are not readily available for \text{DO} data in practice, limiting 
the applicability of this approach. Technical details of these approaches and 
their formal connections are provided in Supplementary 
Material~\ref{supp:do} and \ref{supp:connections}.

No existing method accounts for false positive errors in \text{DO} data, and those that address false negative errors require separate detection covariates that are rarely available in practice. In 
Section~\ref{sec:modelling}, we propose a novel occupancy-style hierarchical 
model that does precisely this, incorporating the unknown visit probability $q$ 
explicitly into the likelihood and achieving identifiability through prior 
distributions on $p_{11}$ and $p_{10}$ that reflect the natural constraint 
$p_{11} > p_{10}$, established in Section~\ref{bg1.1}.

\subsection{Connections between models}\label{bg1.4}

The modelling approaches reviewed in Sections~\ref{bg1.1}--\ref{bg1.3}, whilst 
appearing methodologically distinct, are formally connected within our unified 
framework. We summarise the key connections here; full technical details are 
provided in Supplementary Material~\ref{supp:connections}.

The connections between Maxent, the logistic regression model, and the IPPP 
have been extensively studied. \cite{renner2013equivalence} showed that Maxent 
\citep{phillips2004maximum} and the IPPP \citep{warton2010poisson} are 
equivalent, differing only in the intercept due to scale dependence in Maxent 
when the study area is divided into grid cells. \cite{fithian2012finite} further 
demonstrated that Maxent and the IPPP lead to the same probability density, and 
that the logistic regression model and the IPPP are asymptotically equivalent 
with large background samples. The key practical implication is that 
pseudo-absences in the logistic regression model, quadrature points in the IPPP, 
and background data in Maxent all serve the same purpose: to represent the 
distribution of environmental conditions across $\mathcal{A}$ and to approximate 
the integral over the study area in the likelihood. \cite{barbet2012selecting} 
recommended using at least 10,000 background samples for stable performance, 
consistent with the default in Maxent \citep{elith2011statistical}.

A further connection, not explicitly recognised, is between the \text{DO} models 
reviewed in Section~\ref{bg1.3} and the \text{DN-1} model of 
Section~\ref{bg1.2}. The observation model underlying \text{DO} data is 
identical to that of \text{DN-1} --- both involve a single visit, the same 
false positive and false negative error structure, and the same logistic model 
for $\psi_i$ --- but with two critical differences: in \text{DO} data, $q$ is 
unknown and non-detections are not recorded. As a consequence, $p_{11}$ and 
$p_{10}$ are confounded with $q$ in \text{DO} data, so that only the products 
$\theta_{11} = qp_{11}$ and $\theta_{10} = qp_{10}$ are identifiable from the 
data. This confounding is not recognised by any of the existing \text{DO} 
approaches reviewed in Section~\ref{bg1.3}, all of which implicitly set $q = 1$ 
or ignore the visit process entirely. Recognising this connection is the central 
insight of the proposed model in Section~\ref{sec:modelling}: by treating 
\text{DO} data as a noisy, partially observed version of \text{DN-1} data and 
incorporating $q$ explicitly into the likelihood, we obtain a principled 
hierarchical model for \text{DO} data that accounts for both false positive 
and false negative errors without requiring detection covariates.

Although existing \text{DO} approaches provide consistent estimates of the 
slope parameters $\boldsymbol{\beta}$ under certain conditions 
\citep{warton2010poisson, fithian2012finite}, this consistency breaks down in 
the presence of observation error \citep{dorazio2012predicting}. Specifically, 
\cite{dorazio2012predicting} showed that consistent estimation requires 
disjoint covariate sets for the occurrence and detection processes --- a 
condition that is rarely satisfiable in practice for \text{DO} data. Moreover, 
the well-known divergence of the intercept term as background sample size grows 
\citep{owen2007infinitely, renner2013equivalence} is a consequence of ignoring 
the observation error structure: as we show in Section~\ref{sec:simstudy}, this 
divergence disappears under our proposed model, which correctly accounts for 
observation error within the unified hierarchical framework.
\section{Detection-only hierarchical model }\label{sec:modelling}

As established in Section~\ref{bg1.4}, the observation process underlying \text{DO} 
data is structurally analogous to that of \text{DN-1}: both involve a single visit, 
the same false positive and false negative error structure, and the same logistic 
model for $\psi_i$. The critical differences are that in \text{DO} data the 
probability of site visit $q$ is unknown and non-detections are not recorded. By explicitly modelling $q$, our proposed model naturally absorbs all processes that 
determine whether a site generates a detection record --- including recorder effort, 
reporting behaviour, and habitat accessibility. This means that $q$ should be interpreted as a 
composite quantity rather than a pure visit probability, and the products 
$\theta_{11} = qp_{11}$ and $\theta_{10} = qp_{10}$ reflect this composite nature. This 
section formalises this analogy, derives the likelihood for \text{DO} data, studies 
identifiability, and specifies the prior distributions and posterior used for 
inference. We refer to the proposed model hereafter as the detection-only model (DO-M).

\subsection{Observation model and likelihood}\label{sec:likelihood}

From Figure~\ref{fig:tree}, the conditional distribution of the observed data $Y_i$ 
given the latent occupancy state $Z_i$ and visit indicator $V_i$ is
\begin{equation}
    \mathbb{P}(Y_i = y \mid Z_i = z, V_i = v) =
    \begin{cases}
    q\,p_{11}, & y=1, \; z=1, \; v=1,\\
    q\,(1-p_{11}), & y=\text{NA}, \; z=1, \; v=1,\\
    q\,p_{10}, & y=1, \; z=0, \; v=1,\\
    q\,(1-p_{10}), & y=\text{NA}, \; z=0, \; v=1,\\
    (1-q), & y=\text{NA}, \; z=1, \; v=0,\\
    (1-q), & y=\text{NA}, \; z=0, \; v=0.
    \end{cases}
    \label{eq:obsmodel}
\end{equation}
Since $Z_i$ and $V_i$ are independent, applying the total probability theorem and 
marginalising over $Z_i$ gives
\begin{equation}
\begin{aligned}
\mathbb{P}(Y_i = 1)
&= \mathbb{P}(Z_i = 1)\,\mathbb{P}(V_i = 1)\,\mathbb{P}(Y_i = 1 \mid Z_i = 1) \\
&\quad + \mathbb{P}(Z_i = 0)\,\mathbb{P}(V_i = 1)\,\mathbb{P}(Y_i = 1 \mid Z_i = 0) \\
&= qp_{11}\,\psi_i + qp_{10}\,(1 - \psi_i).
\end{aligned}
\label{eq:prob1}
\end{equation}
Equation~\ref{eq:prob1} reveals that $p_{11}$ and $p_{10}$ are confounded with $q$ 
in \text{DO} data: only the products $\theta_{11} = qp_{11}$ and 
$\theta_{10} = qp_{10}$ are estimable from the data alone. Reparameterising in 
terms of $\theta_{11}$ and $\theta_{10}$, 
\begin{equation}
\mathbb{P}(Y_i = 1) = \theta_{11}\,\psi_i + \theta_{10}\,(1 - \psi_i),
\label{eq:prob2}
\end{equation}
so that the observation model is
\begin{equation}
Y_i \mid \psi_i, \theta_{11}, \theta_{10} \sim 
\text{Bernoulli}\!\left(\theta_{11}\,\psi_i + \theta_{10}\,(1 - \psi_i)\right).
\label{eq:obsmod2}
\end{equation}
Note that setting $p_{10} = 0$, as in the single-visit model of 
\cite{lele2012dealing}, implies $\theta_{10} = 0$ and 
Equation~\ref{eq:prob2} reduces to 
$\mathbb{P}(Y_i = 1) = \theta_{11}\psi_i$, showing that our model 
nests this as a special case. Comparing Equation~\ref{eq:prob2} with the \text{DN-1} likelihood in 
Equation~\ref{eq:dn1lik}, the observation model is identical in form, 
with $\theta_{11} = qp_{11}$ and $\theta_{10} = qp_{10}$ in place of 
$p_{11}$ and $p_{10}$, reflecting the absorption of the unknown visit 
probability $q$ into the detection probabilities. Thus, the interpretation \(\theta_{11}\) and \(\theta_{10}\) is not straightforward as it combines detection probabilities and sampling bias (i.e., recorder interest, habitat accessibility, etc.). 

Assuming conditional independence of $Y_i$ across sites given $\theta_{11}$ and 
$\theta_{10}$, the likelihood for \text{DO} data is
\begin{equation}
\mathcal{L}(\boldsymbol{\beta},\,\theta_{11},\,\theta_{10})
= \prod_{i=1}^{n} 
\Big(\theta_{11}\psi_i + \theta_{10}(1-\psi_i)\Big)^{Y_i} 
\Big(1 - \theta_{11}\psi_i - \theta_{10}(1-\psi_i)\Big)^{1-Y_i},
\label{eq:lik}
\end{equation}
with log-likelihood
\begin{equation}
\begin{aligned}
\ell(\boldsymbol{\eta},\,\theta_{11},\,\theta_{10}) 
= &\sum_{i:\,Y_i=1}
\log\!\left(
\frac{\theta_{10} + \theta_{11}\,e^{\eta_i(\mathbf{x})}}
{1 + e^{\eta_i(\mathbf{x})}}
\right) \\
&+ \sum_{i:\,Y_i=0}
\log\!\left(
\frac{(1-\theta_{10}) + (1-\theta_{11})\,e^{\eta_i(\mathbf{x})}}
{1 + e^{\eta_i(\mathbf{x})}}
\right).
\end{aligned}
\label{eq:loglik}
\end{equation}
Equation~\ref{eq:loglik} suffices for \text{DO} data with well-defined sites and 
fixed $n$, where the $n - n_1$ sites with $Y_i = \text{NA}$ contribute to the 
second sum with $Y_i$ coded as 0, interpreted as the absence of a detection record 
rather than a verified non-detection. When $n$ is unknown, as in the case of 
incidentally collected records, background samples drawn uniformly from 
$\mathcal{A}$ are used to approximate the integral over the study area. Following 
\cite{owen2007infinitely}, replacing the second sum in Equation~\ref{eq:loglik} 
with an integral over $\mathcal{A}$ gives
\begin{equation}
\begin{aligned}
\ell(\boldsymbol{\eta},\,\theta_{11},\,\theta_{10}) 
= &\sum_{i=1}^{n_1}
\log\!\left(
\frac{\theta_{10} + \theta_{11}\,e^{\eta_i(\mathbf{x})}}
{1 + e^{\eta_i(\mathbf{x})}}
\right) \\
&+ n_0 \int_{\mathcal{A}}
\log\!\left(
\frac{(1-\theta_{10}) + (1-\theta_{11})\,e^{\eta_i(\mathbf{x})}}
{1 + e^{\eta_i(\mathbf{x})}}
\right)
\pi_{\mathcal{A}}(s)\,ds,
\end{aligned}
\label{eq:loglik2}
\end{equation}
where $\int_{\mathcal{A}}(\cdot)\pi_{\mathcal{A}}(s)\,ds$ approximates 
$\mathbb{P}(Y_i = 0 \mid \mathbf{x})$ across $\mathcal{A}$, with the approximation 
improving as $n_0$ grows. Replacing $\pi_{\mathcal{A}}(s)$ with the uniform 
distribution $1/n_0$ and evaluating the integral over the $n_0$ background samples 
yields
\begin{equation}
\ell(\boldsymbol{\eta},\,\theta_{11},\,\theta_{10}) 
= \sum_{i=1}^{n_1}
\log\!\Big(\mathbb{P}(Y_i = 1 \mid \mathbf{x})\Big)
+ \sum_{i=n_1+1}^{n_0}
\log\!\Big(\mathbb{P}(Y_i = 0 \mid \mathbf{x})\Big),
\label{eq:loglik3}
\end{equation}
which has the same form as a standard logistic regression log-likelihood but with 
the observation error structure of $\theta_{11}$ and $\theta_{10}$ embedded in 
$\mathbb{P}(Y_i = 1 \mid \mathbf{x})$ and $\mathbb{P}(Y_i = 0 \mid \mathbf{x})$.
We note that the pseudo-absences used in the second sum are not treated as true 
absences: they are background samples drawn from across $\mathcal{A}$ that may 
contain sites where the species is present, and the false negative error structure 
is accounted for through $\theta_{11}$ and $\theta_{10}$ in the likelihood. This 
distinguishes our approach from naive logistic regression, which treats background 
samples as true absences and ignores observation error entirely.

\subsection{Identifiability}\label{sec:identifiability}

The likelihood in Equation~\ref{eq:lik} exhibits the same symmetry as the 
\text{DN-T} likelihood established in Section~\ref{sec:modelling}: swapping 
$(\theta_{11}, \theta_{10}, \beta)$ and $(\theta_{10}, \theta_{11}, -\beta)$ yields 
an identical likelihood value, resulting in a bimodal posterior with two solutions. 
The ecologically meaningful solution is selected by imposing the constraint 
$p_{11} > p_{10}$ through informative prior distributions on $p_{11}$ and $p_{10}$: 
since a species is more likely to be recorded where it is present than where it is 
absent, the mode corresponding to $\theta_{11} > \theta_{10}$ is the correct one, 
and the mode corresponding to $\theta_{11} < \theta_{10}$ --- in which the sign of 
$\beta$ would be reversed --- is ruled out. This constraint is ecologically 
justifiable and ensures that the coefficients $\beta$ retain their correct ecological 
interpretation, playing the same role here as in the \text{DN-T} model of 
\citet{royle2006generalized}. Whilst separate detection covariates could in principle be used 
to model $p_{11}$ and $p_{10}$, such covariates are rarely available for \text{DO} 
data in practice \citep{moreira2022analysis}, and our approach does not require them. When 
detection covariates are available, the prior construction of \citet{griffin2020modelling} 
can be adopted to ensure that the constraint $p_{11} > p_{10}$ is maintained even 
when both probabilities are modelled as functions of covariates.

\subsection{Prior distributions and posterior}\label{sec:priors}

The observation model in Equation~\ref{eq:obsmod2} is parameterised in terms of 
$\theta_{11} = qp_{11}$ and $\theta_{10} = qp_{10}$, but it is more natural to 
specify prior distributions on $p_{11}$, $p_{10}$, and $q$ separately, since each 
has a direct ecological interpretation: $p_{11}$ is the probability of correctly 
recording a species where it is present, $p_{10}$ is the probability of falsely 
recording it where it is absent, and $q$ is the probability that a site generates 
a detection record, absorbing visit probability, reporting behaviour, and recorder 
interest. The implied prior distributions on $\theta_{11}$ and $\theta_{10}$ are 
then obtained via the change of variable theorem \citep{siegrist2024}.

Let $\theta = qp$, with $q$ and $p$ independent random variables on $(0,1)$ with 
probability density functions $\pi_q$ and $\pi_p$ respectively. The inverse 
transformation is $(q = q,\, p = \theta/q)$, and since 
$\partial\theta/\partial p = q$, the Jacobian is $|\partial\theta/\partial p|^{-1} 
= 1/q$. Since $q \in (0,1)$ and $0 < \theta/q < 1$, we have $0 < \theta < q < 1$ 
for any given $\theta$, so the implied prior distribution on $\theta$ is
\begin{equation}
\pi_{\theta}(\theta) = \int_{\theta}^{1}
\pi_q(q)\, \pi_p\!\left(\frac{\theta}{q}\right) \frac{1}{q} \, dq, 
\quad 0 < \theta < 1.
\label{eq:cov}
\end{equation}
Applying this result separately to $\theta_{11} = qp_{11}$ and 
$\theta_{10} = qp_{10}$, the marginal prior distributions are
\begin{equation}
\pi_{\theta_{11}}(\theta_{11}) = \int_{\theta_{11}}^{1}
\pi_q(q)\, \pi_{p_{11}}\!\left(\frac{\theta_{11}}{q}\right) \frac{1}{q} \, dq, 
\qquad 
\pi_{\theta_{10}}(\theta_{10}) = \int_{\theta_{10}}^{1}
\pi_q(q)\, \pi_{p_{10}}\!\left(\frac{\theta_{10}}{q}\right) \frac{1}{q} \, dq.
\label{eq:margpriors}
\end{equation}
Although these induced prior distributions do not admit closed-form expressions 
for general Beta hyperparameters, they can be characterised numerically via 
Equation~\ref{eq:margpriors} for any given choice of hyperparameters. The induced priors for $\theta_{11}$ and $\theta_{10}$ under different prior specification for $p_{11}$, $p_{10}$ and $q$ are presented in Figure~\ref{fig:induced-priors} of Supplementary Material~\ref{supp:induced-priors}.

Prior distributions on $p_{11}$ and $p_{10}$ are specified as Beta distributions, 
$p_{11} \sim \text{Beta}(a_{11}, b_{11})$ and 
$p_{10} \sim \text{Beta}(a_{10}, b_{10})$, with hyperparameters chosen to reflect 
the constraint $p_{11} > p_{10}$ and any available empirical knowledge about 
observer accuracy \citep{groom2017characterisation}. A uniform prior distribution 
$q \sim \text{Beta}(1, 1)$ is specified for $q$, reflecting genuine uncertainty 
about this composite quantity: since visit probability, reporting behaviour, and 
recorder interest are inseparable in \text{DO} data, informative prior 
specification would require external knowledge of recorder effort or survey 
intensity that is rarely available in practice. Prior distributions on 
$\boldsymbol{\beta}$ are chosen following \cite{newman2025logistic} to induce a 
meaningful prior distribution on $\psi_i$.

Combining the log-likelihood in Equation~\ref{eq:loglik} with the prior 
distributions on $q$, $p_{11}$, $p_{10}$, and $\boldsymbol{\beta}$, the log 
posterior distribution is
\begin{equation}
\begin{aligned}
\log \pi(\boldsymbol{\beta}, q, p_{11}, p_{10} \mid \mathbf{Y}, \mathbf{x}) 
\propto\; &\sum_{i=1}^{n} \Bigl[ 
Y_i \log\!\big(\theta_{11}\psi_i + \theta_{10}(1-\psi_i)\big) \\
&+ (1-Y_i) \log\!\big(1 - \theta_{11}\psi_i - \theta_{10}(1-\psi_i)\big)\Bigr] \\
&+ \sum_{k=1}^{p} \log \pi(\beta_k) \\
&+ \log \pi_q(q) 
+ \log \pi_{p_{11}}(p_{11})
+ \log \pi_{p_{10}}(p_{10}),
\end{aligned}
\label{eq:posterior}
\end{equation}
where $\theta_{11} = qp_{11}$ and $\theta_{10} = qp_{10}$ are deterministic 
functions of $q$, $p_{11}$, and $p_{10}$. The posterior distribution in 
Equation~\ref{eq:posterior} is not available in closed form. Samples are obtained 
using the Hamiltonian Monte Carlo algorithm \citep{neal2011}, implemented via the 
\texttt{cmdstanr} package \citep{cmdstan} in \texttt{R}, with $\theta_{11}$ and 
$\theta_{10}$ defined as deterministic functions in the model. This approach is 
supported natively by Stan \citep{cmdstan} and other Bayesian software such as 
NIMBLE \citep{deValpine2017}, which handle deterministic relationships between 
parameters without requiring a closed-form expression for the induced prior on 
$\theta_{11}$ and $\theta_{10}$. Since only $\theta_{11}$ and $\theta_{10}$ enter 
the likelihood, $q$, $p_{11}$, and $p_{10}$ are not separately identifiable from 
the data; their marginal posteriors are informed by the data only through the 
products $\theta_{11}$ and $\theta_{10}$, and should not be interpreted as 
individually meaningful posterior quantities. Convergence is assessed using the 
$\hat{R}$ statistic \citep{gelman1992} and effective sample size diagnostics.
\section{Simulation study}\label{sec:simstudy}

In this section, we perform an extensive simulation study to evaluate the 
performance of \text{DO-M} across different levels of site visit probability 
$q$, and to compare its performance in estimating the coefficients of 
environmental conditions on occurrence probability against that of 
\text{DN-}$T$ models. We simulate $n = 5{,}000$ sites and perform 500 
simulation replicates for each model considered.

Throughout the simulation study, five models are compared. The oracle model 
(\text{Oracle}) is fitted to the true latent occurrence states $Z_i$, which 
are known in a simulation setting but never observable in practice; it 
therefore represents an upper bound on estimation performance that no real 
model can attain, and serves as a reference against which all other models 
are assessed. The generalised linear model (\text{GLM}) represents the naive 
approach commonly used in practice: it is fitted to \text{DN-1} data treating 
every detection as a confirmed presence and every non-detection as a confirmed 
absence, implicitly assuming $p_{11} = 1$ and $p_{10} = 0$, and ignoring 
observation error entirely. The single-visit and two-visit 
detection/non-detection models (\text{DN-1-M} and \text{DN-2-M}) are fitted 
to \text{DN-1} and \text{DN-2} data respectively, both accounting for 
observation error; \text{DN-2-M} represents the richer data setting in which 
repeated visits provide additional information about the observation process, 
and serves as a reference for what is gained by moving from a single to two 
visits. Finally, \text{DO-M} is fitted to \text{DO} data as described in 
Section~\ref{sec:modelling}. Although \text{DO-M} and \text{DN-1-M} share 
the same underlying observation model, they differ fundamentally in the data 
available: \text{DN-1-M} observes both detections and non-detections from a 
planned visit, whereas \text{DO-M} observes only detections from visits of 
unknown probability $q$.

Three environmental covariates are used throughout: $x_1$ and $x_3$ are 
continuous, simulated from a standard normal distribution, and $x_2$ is a 
binary covariate representing a categorical environmental condition. 
Section~\ref{subsec:simstudy1} investigates estimation of the coefficients 
of environmental conditions under correct model specification across 
different values of $q$ and species prevalence scenarios. 
Section~\ref{subsec:simstudy2} investigates robustness to covariate 
misspecification, considering three cases: omission of an important 
covariate, substitution of an important covariate with an uninformative 
one, and inclusion of an irrelevant covariate. 
Section~\ref{subsec:simstudy3} examines the behaviour of the intercept 
and coefficients as background sample size grows. 
Section~\ref{subsec:simstudy4} compares the predictive performance of 
\text{DO-M} against machine learning and other benchmark models.

\subsection{Estimation of coefficients of environmental 
            conditions}\label{subsec:simstudy1}

Prior distributions for $\boldsymbol{\beta}$ are chosen following 
\cite{newman2025logistic} to induce a meaningful prior distribution on 
$\psi_i$, with the intercept $\beta_0$ treated separately to reflect 
uncertainty about baseline species prevalence. The hyperparameters of 
the Beta prior distributions for $p_{11}$ and $p_{10}$ are chosen to 
reflect realistic observer accuracy in species occurrence surveys: the 
Beta$(8, 2)$ prior distribution places most mass above $0.5$ for 
$p_{11}$, reflecting a reasonably high probability of correctly recording 
a species where it is present, whilst the Beta$(1, 19)$ prior distribution 
places most mass close to zero for $p_{10}$, reflecting a low probability 
of falsely recording a species where it is absent. Together these prior 
distributions enforce the identifiability constraint $p_{11} > p_{10}$ 
established in Section~\ref{sec:identifiability}. A uniform prior 
distribution $q \sim \text{Beta}(1, 1)$ is specified for $q$, reflecting 
genuine uncertainty about this composite quantity, as discussed in 
Section~\ref{sec:priors}. The prior distributions for $p_{11}$ and $p_{10}$ and the induced priors for $\theta_{11}$ and $\theta_{10}$ are shown in the Figure~\ref{fig:sim-study-priors} in Supplementary Material~\ref{supp:induced-priors}.

Specifically, the prior distributions used are
\begin{align}
\beta_0 &\sim \text{N}(0, 4), \nonumber \\
\beta_k &\sim \text{N}(0, 1), \quad k = 1, \ldots, p, \nonumber \\
p_{11} &\sim \text{Beta}(8, 2), \nonumber \\
p_{10} &\sim \text{Beta}(1, 19), \nonumber \\
q &\sim \text{Beta}(1, 1). \nonumber
\end{align}

Detection-only data are simulated for six levels of the site visit 
probability $q \in \{0.5, \allowbreak 0.6, \allowbreak 0.7, \allowbreak 
0.8, \allowbreak 0.9, \allowbreak 1.0\}$ to investigate the influence of 
$q$ on the intercept and coefficients of environmental conditions. The 
observation probabilities are fixed at $(p_{11}, p_{10}) = (0.80, 0.05)$ 
throughout. The simulation is conducted under three species prevalence 
scenarios, determined by the value of the intercept $\beta_0$, with the 
coefficients of environmental conditions fixed at $(\beta_1, \beta_2, 
\beta_3) = (1.2, 1.0, -1.0)$ throughout:
\begin{itemize}
    \item Scenario 1: $\beta_0 = 0$, giving a baseline occurrence 
    probability of $\bar{\psi} = 0.50$.
    \item Scenario 2: $\beta_0 = 1.5$, giving a baseline occurrence 
    probability of $\bar{\psi} = 0.82$.
    \item Scenario 3: $\beta_0 = -1.5$, giving a baseline occurrence 
    probability of $\bar{\psi} = 0.18$.
\end{itemize}
The performance of each model is assessed using three metrics computed 
across the 500 simulation replicates: the mean relative bias (RB), the 
mean root mean squared error (RMSE), and the coverage (COV) of the 95\% 
posterior credible intervals. We note that RB is undefined when the true 
parameter value is zero, and is therefore not reported for $\beta_0$ 
under Scenario 1.

Table~\ref{tab:sim1} in Supplementary 
Material~\ref{supp:simstudy} presents results for all five models under 
the three prevalence scenarios, with DO-M results shown for $q = 1$ to 
facilitate direct comparison with DN-1-M, DN-2-M, Oracle, and GLM, all 
of which implicitly assume all sites are visited. The DN-2-M has the 
lowest RB and RMSE across all scenarios, as expected since multiple visits 
provide more information about the observation process. The Oracle and 
DN-2-M achieve comparable RB and coverage, with DN-2-M showing slightly 
higher RMSE. DO-M achieves RB, RMSE, and coverage comparable to DN-1-M 
across all three prevalence scenarios, demonstrating that the proposed 
model provides unbiased estimates of the coefficients of environmental 
conditions from \text{DO} data when observation error is properly 
accounted for. The GLM performs poorly across all scenarios: RB and RMSE 
are far from zero, and coverage collapses to zero, meaning that for none 
of the 500 simulation replicates does the 95\% posterior credible interval 
contain the true parameter value. This confirms that ignoring observation 
error in \text{DO} data leads to severely biased inference on the 
coefficients of environmental conditions, as established in 
Section~\ref{sec:intro}.

Table~\ref{tab:sim2} in Supplementary 
Material~\ref{supp:simstudy} presents results for DO-M across all six 
values of $q$ under the three prevalence scenarios. As $q$ increases 
towards 1, RMSE decreases, reflecting lower uncertainty when more sites 
are visited. The difference in RB across values of $q$ is small --- for 
example, comparing $q = 0.5$ and $q = 1$, the RB values are comparable 
--- but the difference in RMSE is more pronounced, with higher uncertainty 
when $q = 0.5$. Coverage of the 95\% posterior credible intervals is 
comparable across all values of $q$ and all three prevalence scenarios, 
confirming that DO-M provides well-calibrated uncertainty quantification 
regardless of the visit probability. DO-M outperforms GLM for all values 
of $q$ considered.

\subsection{Effect of model misspecification on coefficient 
            estimation}\label{subsec:simstudy2}

We investigate the robustness of each model to three types of model 
misspecification that are likely to arise in practice. Results for DO-M 
are shown for $q = 1$ to isolate the effect of misspecification from the 
additional uncertainty introduced by unknown $q$. Three cases are 
considered:
\begin{itemize}
    \item Case 1: an important covariate $x_3$ is omitted from the model 
    for $\psi_i$.
    \item Case 2: an important covariate $x_3$ is replaced by $x_3^*$, 
    a covariate uncorrelated with $x_3$ and carrying no information about 
    occurrence.
    \item Case 3: an irrelevant covariate $x_4$, with true coefficient 
    zero, is included in the model for $\psi_i$.
\end{itemize}

Table~\ref{tab:sim3} in Supplementary 
Material~\ref{supp:simstudy} presents results for all five models under 
the three misspecification cases. Under Cases 1 and 2, all models yield 
biased estimates of $\beta_1$ and $\beta_2$, with high RB and high RMSE, 
regardless of whether observation error is accounted for. This confirms 
that model misspecification in the form of omitted or incorrect covariates 
has consequences for all models, not just \text{DO-M}, and underscores 
the importance of careful covariate selection. The performance of DO-M 
under Cases 1 and 2 is comparable to that of DN-1-M. Under Case 3, all 
models perform well in terms of RB and coverage, confirming that the 
inclusion of an irrelevant covariate does not introduce bias, though RMSE 
is slightly elevated relative to the correctly specified model. The 
performance of DO-M under Case 3 is comparable to that of DN-1-M.

\subsection{Effect of background sample size on coefficient and intercept 
            estimation}\label{subsec:simstudy3}

We investigate the behaviour of the intercept and coefficients as the 
number of background samples grows, following the infinitely imbalanced 
logistic regression framework of \cite{owen2007infinitely}. The number 
of detections $n_1$ is held fixed whilst the number of background samples 
$n_0$ is increased up to $140{,}000$. Results for all coefficients and 
the intercept are shown in Figure~\ref{fig:convergence}; additional 
results for further prevalence scenarios are provided in Supplementary 
Material~\ref{supp:convergence}.

For the GLM, the coefficient estimates are attenuated due to 
contamination in both the detections and the background samples 
\citep{dorazio2012predicting}, converging to a non-zero but biased value 
as $n_0 \to \infty$. The intercept diverges to $-\infty$ as $n_0$ grows, 
consistent with the results of \cite{owen2007infinitely} and 
\cite{renner2013equivalence} for naive logistic regression in the absence 
of observation error; \cite{renner2013equivalence} attribute this 
divergence to model misspecification. In contrast, DO-M yields convergent 
estimates of both the coefficients and the intercept to their true values 
as $n_0$ grows, with coefficients converging faster than the intercept. 
The intercept converges more rapidly as $n_1$ increases, reflecting the 
benefit of additional detection records. This convergence is a direct 
consequence of the hierarchical structure of DO-M, which correctly 
accounts for observation error and incorporates informative prior 
distributions on $p_{11}$ and $p_{10}$.
\begin{figure}[htbp]
  \centering
  \includegraphics[width=\linewidth]{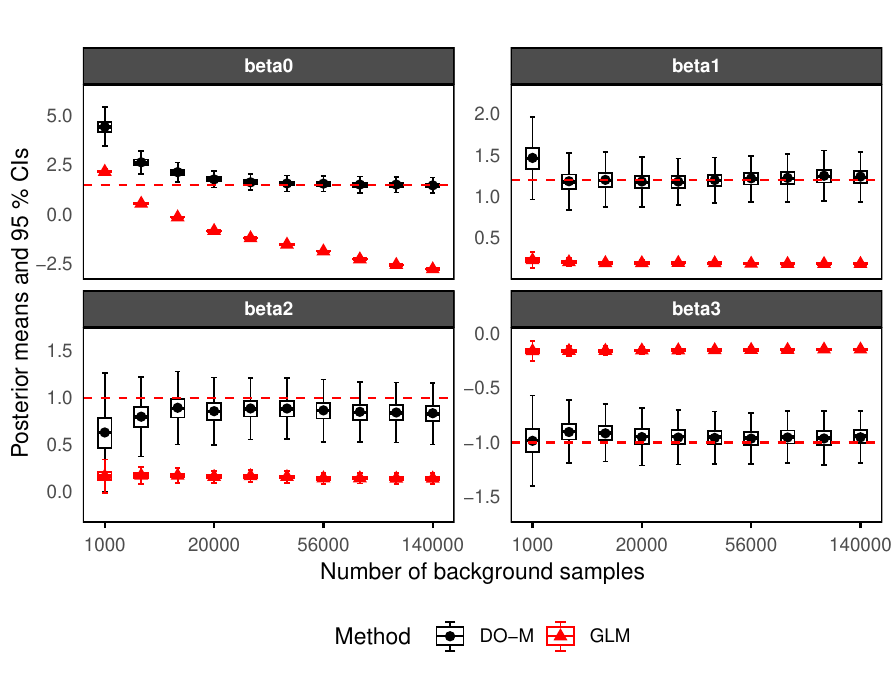}
  \caption{For both the generalised linear model (GLM) and the detection-only model (DO-M) the posterior means and credible intervals of intercept ($\beta_0$), and coefficients ($\beta_1$, $\beta_2$, and $\beta_3$) are presented with $\beta_0 = 1.5$ corresponding to a baseline occurrence probability 0.82 ($\bar{\psi = 0.82}$) as the number of background samples 
  $n_0$ increases, with the number of detections $n_1$ held fixed. The 
  GLM coefficient estimates converge to a non-zero but biased value as 
  $n_0 \to \infty$, whilst the intercept diverges to $-\infty$, 
  consistent with the results of \cite{owen2007infinitely} and 
  \cite{renner2013equivalence}. In contrast, DO-M yields convergent 
  estimates of both the coefficient and the intercept to their true 
  values (shown by the dashed red line) as $n_0$ grows, a direct 
  consequence of correctly accounting for observation error within the 
  hierarchical framework. The two other scenarios where $\beta_0 = 0$ and $\beta_0 = -1.5$ for baseline occurrence probability of $\bar{\psi} = 0.5$ and $\bar{\psi} = 0.18$ are presented in Figure~\ref{fig:supp_convergence_beta0} and \ref{fig:supp_convergence_beta2}  of the Supplemetary Material~\ref{supp:convergence}.}
  \label{fig:convergence}
\end{figure}
\subsection{Predictive performance}\label{subsec:simstudy4}

We compare the predictive performance of DO-M against four benchmark 
models: the GLM, Maximum entropy (Maxent, implemented via the 
\texttt{maxnet} package in \texttt{R}), boosted regression trees (BRT), 
and random forest (RF). Detection-only data are simulated for $q = 0.5$ 
and $q = 1$ with a total sample size of $15{,}000$ sites. A $K$-fold 
cross-validation with $K = 5$ is performed, using $10{,}000$ sites for 
training and $5{,}000$ for testing. We perform 500 simulation replicates 
to obtain stable estimates of predictive performance metrics; the averages 
of the area under the receiver operating characteristic curve (AUC) and 
the Brier score (BS) across the 500 replicates are reported in 
Table~\ref{tab:sim4} of Supplementary 
Material~\ref{supp:simstudy}. The AUC and BS are computed against the 
true latent occurrence states $Z_i$ in the test data, so that all models 
are evaluated on their ability to recover the true underlying distribution 
of species occurrence rather than the noisy observed detections.

DO-M achieves AUC values comparable to all benchmark models for both 
$q = 0.5$ and $q = 1$, including the GLM which ignores observation error 
entirely. This is expected: AUC measures the ability to discriminate 
between occupied and unoccupied sites, and a model can discriminate well 
even when its coefficient estimates are severely biased, as demonstrated 
in Section~\ref{subsec:simstudy1}. The BS, which measures 
calibration of predicted occurrence probabilities rather than 
discrimination, is the more informative metric here. When $q = 1$, the 
BS of DO-M is comparable to all other models, reflecting well-calibrated 
predictive uncertainty when all sites are visited. When $q = 0.5$, the 
BS of DO-M is substantially higher than that of the benchmark models, 
reflecting the greater uncertainty in predicted occurrence probabilities 
when only half of sites are visited and $q$ is unknown. This higher BS 
is a direct consequence of DO-M correctly propagating the genuine 
uncertainty in the data through the posterior distribution --- uncertainty 
that the benchmark models fail to capture by ignoring the observation 
process entirely. A notable advantage of DO-M over all benchmark models 
is that it yields posterior credible intervals for all quantities of 
interest, including predictive performance metrics, which is not possible 
with the non-Bayesian benchmark models. The benchmark models are faster 
computationally than DO-M; the practical implications of this are 
discussed in Section~\ref{sec:discussion}.

\section{Case studies}\label{sec:casestudy}

We present two case studies to illustrate the proposed framework. In 
both case studies, DO-M is applied to \text{DO} data. In Case 
study~1, \text{DN-1} data are also available for the same species and 
sites, providing a unique opportunity to directly compare the results 
of DN-1-M and DO-M and to assess how much is lost by having only 
\text{DO} data rather than the more informative \text{DN-1} data. In 
Case study~2, only \text{DO} data are available, and DO-M is the only 
model that can be applied. In both case studies, prior distributions 
for $p_{11}$ and $p_{10}$ are $p_{11} \sim \mathrm{Beta}(5, 2)$ and $p_{10} \sim \mathrm{Beta}(2, 20)$ respectively, our choice is informed by empirical estimates of 
observer accuracy for vascular plant species 
\citep{groom2017characterisation}, and a uniform prior distribution 
$q \sim \text{Beta}(1,1)$ is specified for $q$, reflecting genuine 
uncertainty about this composite quantity, the induced priors for $\theta_{11}$ and $\theta_{10}$ are shown in Figure~\ref{fig:case-study-priors} of Supplementary Material~\ref{supp:induced-priors}. Prior distributions for 
$\boldsymbol{\beta}$ are chosen following \cite{newman2025logistic}. 
In both case studies, continuous environmental variables are 
standardised to have mean zero and unit variance prior to model 
fitting. Following \cite{andersen2022incorporation},  for both Case study 1 and 2,
 variables that exhibit high correlation with latitude 
--- for example, the bio-climatic variables Rain, Mas, Sseas, Tseas, Temperature, including Dry and Wet nitrogen deposition  --- are 
latitude-adjusted prior to model fitting, and latitude is included 
as an additional explanatory variable.  We first standardised the latitude to have a mean of zero and unit variance. Using latitude as a predictor or smoothing variable, we fit a Generalised Additive Model (GAM) ($\text{gam}(response) \sim \text{s}(latitude)$) and then extract the residual, free from latitudinal variation. 
The extracted residuals are the latitude-adjusted variables, which we use in our analysis. \cite{andersen2022incorporation} shows that including latitude-adjusted variables improves the predictive performance of Maxent. Also, Figure~\ref{fig:supp_convergence_nz42} and \ref{fig:supp_convergence_CS1} in Supplementary Material~\ref{supp:convergence} show that latitude-adjustment facilitates the limiting behaviour observed in Figure~\ref{fig:convergence} of the Simulation study.
A variable is considered to have a meaningful effect on 
occurrence if its 95\% posterior credible interval excludes zero.

\subsection{Vascular plants from New Zealand}\label{subsec:nz}

The New Zealand vascular plant data are obtained from the National 
Center for Ecological Analysis and Synthesis (NCEAS) working group 
and are available via the \texttt{disdat} package in \texttt{R} 
\citep{elith2020presence}, which contains data on 226 species across 
six geographical regions. We selected New Zealand because it provides 
the largest number of vascular plant observations relative to 
geographical area among the six regions. The data are gathered across 
grids of 100-metre resolution, with a maximum of one record per 
species per grid. There are 3,088 \text{DO} records and 19,020 
\text{DN-1} records, together with 10,000 background samples and 
measurements for thirteen environmental variables, eleven continuous 
and two categorical. Full details of the data collection process and 
environmental variables are provided in \cite{elith2020presence}; 
descriptions and summary statistics for the environmental variables 
used in our analysis are presented in Tables~\ref{tab:variables} and 
\ref{tab:descriptive} in Supplementary 
Material~\ref{supp:casestudy1}. Following \cite{elith2020presence}, 
we excluded \textit{toxicats} due to inconsistent category 
definitions across the \text{DO} and \text{DN-1} datasets, and 
\textit{mat}, \textit{r2pet}, and \textit{vpd} due to collinearity 
with retained variables.

The \texttt{disdat} package provides \text{DN-1} data for 52 
vascular plant species under anonymised species names. We selected 
species \textit{nz42}, with an observed detection rate of 
$\bar{y} = 0.31$, to represent a species of moderate prevalence. 
Both DN-1-M and DO-M are applied to this species. Crucially, both 
models use the same underlying observations: DN-1-M treats the 
non-detections as informative, whilst DO-M treats them as missing 
records, using only the detections and background samples. This 
allows a direct assessment of how much information is lost by having 
only \text{DO} data rather than the more informative \text{DN-1} 
data.

\begin{figure}[H]
  \centering
  \includegraphics[width=.6\linewidth]{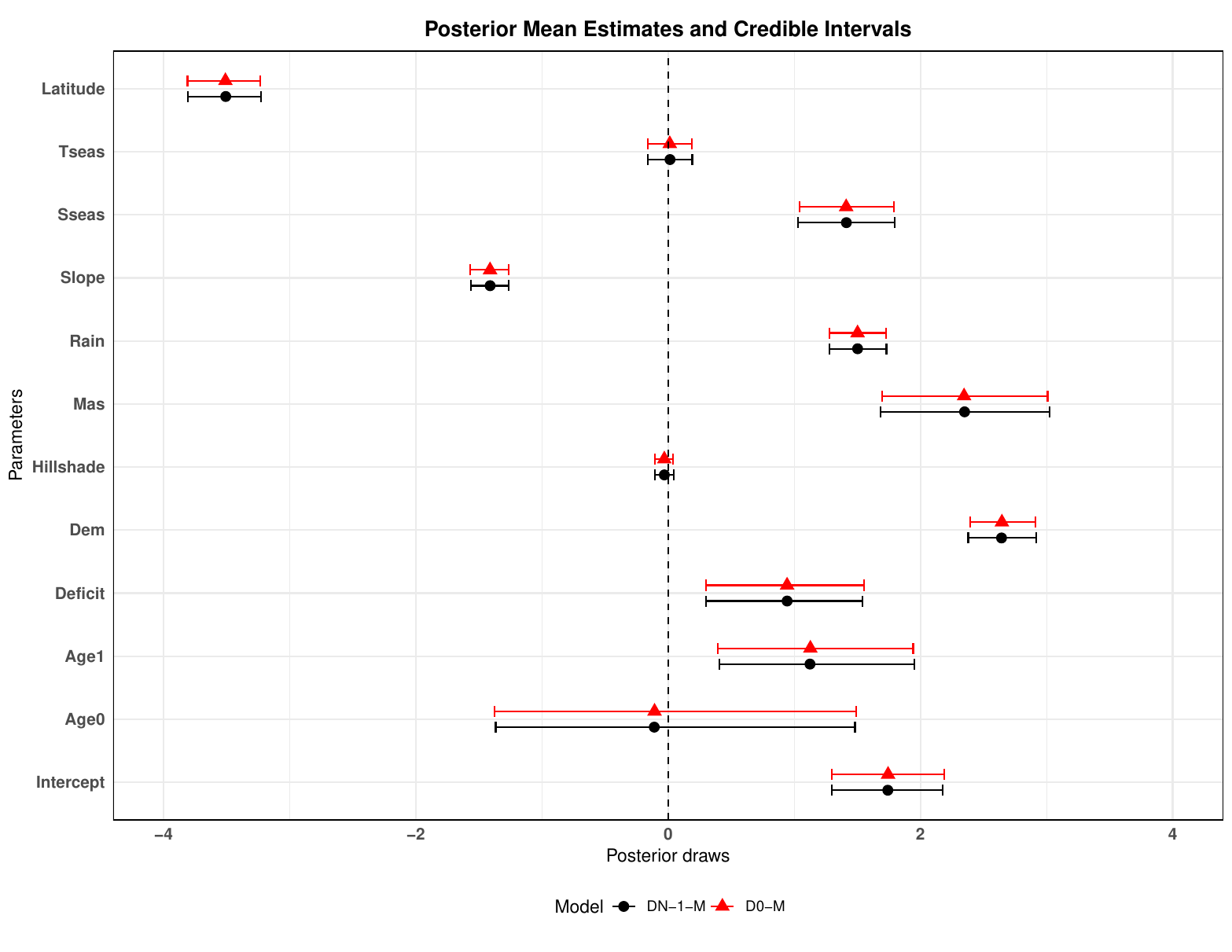}
  \caption{Posterior means and 95\% credible intervals for the 
  intercept and coefficients of environmental conditions from DN-1-M 
  and DO-M fitted to \text{DN-1} and \text{DO} data respectively for 
  species \textit{nz42}. Both models identify the same important 
  variables, with coefficient estimates and credible intervals that 
  are strikingly similar. The vertical dashed line indicates zero; 
  variables whose credible intervals include zero --- Age0, Hillshade, 
  and Tseas --- are not identified as important drivers of occurrence.}
  \label{fig:agreement}
\end{figure}

Results for DN-1-M and DO-M are presented in 
Table~\ref{tab:nz_occ1} in Supplementary 
Material~\ref{supp:casestudy1}, and Figure~\ref{fig:agreement} 
shows the posterior means and credible intervals for both models. 
Both models identify the same set of important variables: Deficit, 
Dem, Mas, Rain, Slope, Sseas, and Tseas. The coefficient estimates 
and credible intervals are strikingly similar across the two models, 
with the DO-M estimates differing from those of DN-1-M by at most 
0.3\% for any coefficient. The close agreement between DN-1-M and 
DO-M demonstrates that DO-M recovers reliable estimates of the 
coefficients of environmental conditions from \text{DO} data alone, 
with no meaningful loss of inferential precision relative to the 
more data-rich DN-1-M.
\subsection{Vascular plants from the United 
Kingdom}\label{subsec:uk}

We apply DO-M to two vascular plant species from the United Kingdom 
using historical \text{DO} data obtained from the Botanical Society 
of Britain and Ireland (BSBI). The BSBI database is described in 
detail by \cite{henniges2022taxonomic} and contains records for more 
than 3,000 plant species at various spatial resolutions, with over 
40 million detection records in total. For our application, we use 
records at 10\,km resolution for the period 2010--2019, with each 
hectad contributing at most one detection record per species. 
Environmental covariates for the same period and resolution, 
including mean rainfall, temperature, land-use types, nitrogen 
deposition, and genome size, are provided by 
\cite{henniges2023changes}. We note that 10\,km resolution is a large area which contains multiple land cover types in a given square. However, the commonest land cover type in each square is used. This introduces ambiguity in the interpretation of the effect sizes for the land cover types. A smaller resolution will be more appropriate for a direct interpretation of effect sizes; however, such data is not readily available across the UK. We proceed to use the 10\,km data and note the ambiguity in the interpretation of the effect sizes. This constitutes a \text{DO} dataset of 
the fixed-$n$ type described in Section~\ref{bg1.3}: the study area 
is discretised into a predetermined grid of $n = 2{,}794$ hectads, 
and the likelihood in Equation~\ref{eq:loglik} is applied directly 
without requiring background samples. Descriptions and summary 
statistics for the environmental variables are presented in 
Tables~\ref{tab:variables2} and \ref{tab:descriptive2} in 
Supplementary Material~\ref{supp:casestudy2}.

We selected two species of contrasting ecology and distribution: 
\textit{Dactylorhiza fuchsii (Druce) Soó} (common spotted orchid) and 
\textit{Carex limosa L.} (bog-sedge). \textit{Dactylorhiza fuchsii}, 
also known as \textit{Dactylorhiza fuchsii}, is a widespread tuberous 
perennial orchid native to the United Kingdom, found across a broad 
range of habitats, including woodland, grassland, and coastal sites, 
with a preference for calcareous or base-rich soils. \textit{Carex 
limosa} is a specialist of nutrient-poor bog and fen habitats, with 
a highly restricted distribution in the United Kingdom concentrated 
in upland areas of Scotland and Ireland. These two species therefore 
provide a contrast between a generalist with a broad environmental 
tolerance and a specialist with narrow habitat requirements, allowing 
us to assess DO-M under different distributional regimes.

\begin{figure}[H]
    \centering
    \begin{subfigure}[t]{0.6\textwidth}
        \centering
        \includegraphics[width=\textwidth]{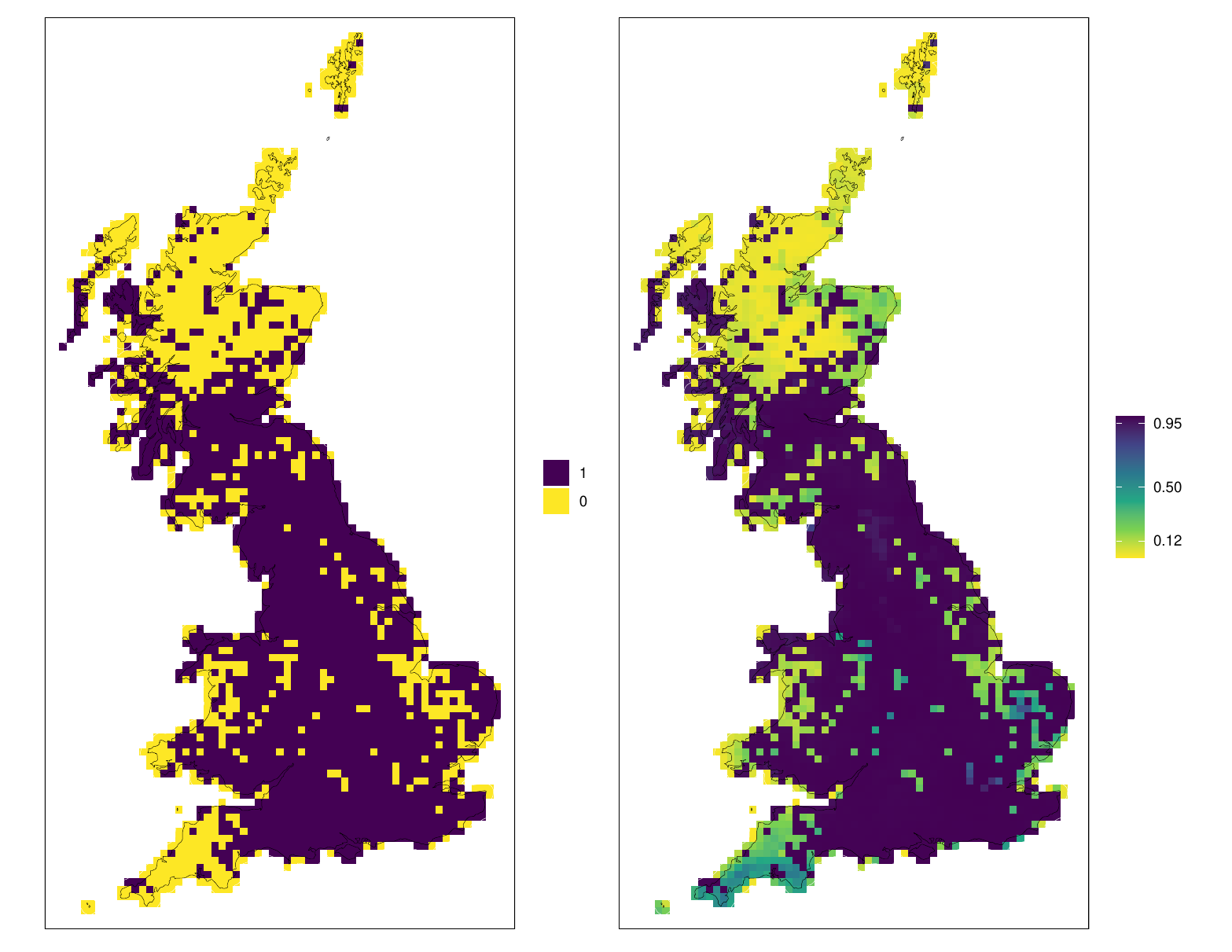}
        \caption{Detection records (left) and predicted occurrence 
        probabilities, i.e., mean posterior predictive probabilities ($\psi_i$) of the latent state $Z_i$ 
        (right) for \textit{Dactylorhiza fuchsii}.}
        \label{fig:OrchisMascula}
    \end{subfigure}
    \hfill
    \begin{subfigure}[t]{0.6\textwidth}
        \centering
        \includegraphics[width=\textwidth]{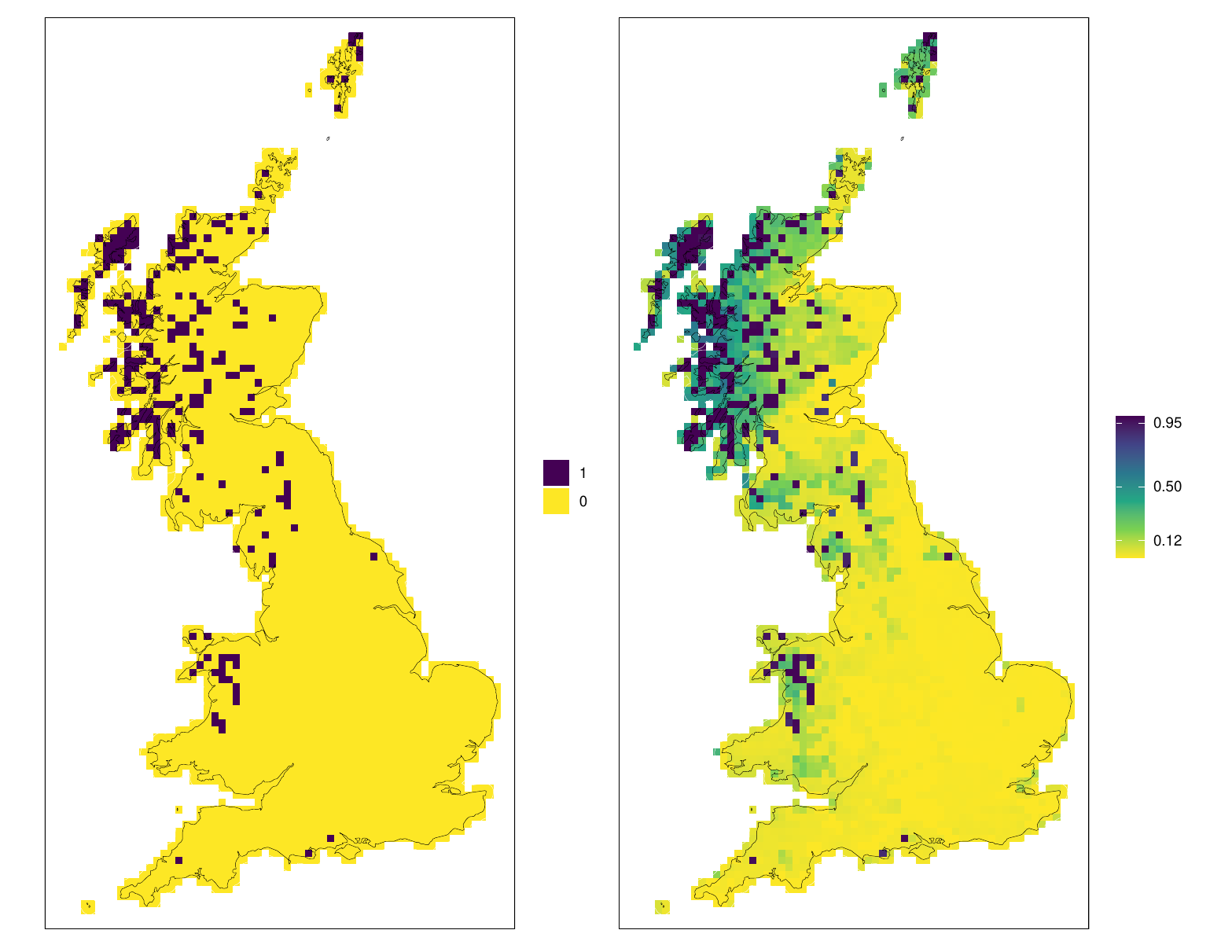}
        \caption{Detection records (left) and predicted occurrence 
        probabilities, i.e., mean of posterior predictive probabilities ($\psi_i$) of the latent state $Z_i$ 
        (right) for \textit{Carex limosa}.}
        \label{fig:CarexLimosa}
    \end{subfigure}
    \caption{Predicted distribution maps for \textit{Dactylorhiza fuchsii} 
    (common spotted orchid) and \textit{Carex limosa} (bog-sedge) 
    from DO-M fitted to BSBI detection records at 10\,km resolution. 
    Predicted occurrence probabilities range from low (yellow) to 
    high (purple). \textit{Orchis fuchsii} is predicted to be 
    widespread across England, Wales, and south of Scotland, whilst 
    \textit{Carex limosa} is predicted to be largely restricted to 
    upland areas of Scotland, consistent with the known distributions 
    of both species.}
    \label{fig:prediction-maps}
\end{figure}

Results are presented in Table~\ref{tab:uk_po1} and the predicted 
distribution maps are shown in Figure~\ref{fig:prediction-maps}. 
For \textit{Dactylorhiza fuchsii}, the model identifies Temperature, Dry 
Nitrogen, Acid and neutral grassland, Improved Grassland, Rocky, 
Arable and horticulture, Urban and suburban, Coniferous woodland, 
and Latitude as important drivers of occurrence. The positive effect 
of temperature is consistent with the species' preference for 
warmer, southern localities. The positive effect of dry nitrogen 
deposition suggests a complex relationship with nitrogen 
availability, whilst the positive effect of improved grassland, acid 
and neutral grasslands, and arable and horticulture reflects the 
species' tolerance of managed agricultural habitats. The negative 
effect of rocky habitat is unsurprising, as this land cover type is 
dominated by exposed, non-vegetated surfaces. The positive effects 
of urban and suburban areas and coniferous woodland are consistent 
with the species' known occurrence along roadsides, railway 
embankments, and other artificial habitats. The negative effect of 
latitude is consistent with the species' distribution, with England, 
Wales, and southern Scotland as its strongholds relative to northern 
Scotland. The estimated observation process parameters 
$\hat{\theta}_{11} = 0.93$ (95\% CI: 0.90, 0.97) and 
$\hat{\theta}_{10} = 0.03$ (95\% CI: 0.003, 0.07) are consistent 
with high recorder interest and reporting rates for this widespread 
and charismatic species, and suggest a small but non-negligible 
false positive rate, reflecting the known difficulty of 
distinguishing \textit{Orchis fuchsii} from other members of the 
genus \textit{Dactylorhiza}.

For \textit{Carex limosa}, the model identifies Temperature, Genome 
size, Bog, Improved grassland, Acid and neutral grassland, Arable 
and horticulture, Sediment coastal, Heath, Coniferous woodland, and 
Latitude as important drivers of occurrence. The positive effect of 
bog is strongly consistent with the species' ecology as a specialist 
of waterlogged, nutrient-poor habitats. The positive effect of 
temperature is consistent with its preference for warm temperatures 
in the growing season. The \textit{Carex} genus (sedge family) is 
well known for having a narrow range of small genome sizes at the lower end of the range reported for 
angiosperms as a whole \citep{lipnerova2013evolution}; the negative effect of 
genome size on occurrence is therefore consistent with the species' 
specialisation to a narrow range of habitats. The negative effects 
of arable and horticulture and improved grassland warrant 
conservation attention, as agricultural intensification tends to 
have a negative effect on the species' distribution. The positive 
effects of acid and neutral grassland and heathland reflect the 
species' preference for acidic soils. The positive effect of 
latitude is consistent with the species' stronghold in 
north-western Scotland and its near-absence from southern England. 
The estimated observation process parameters $\hat{\theta}_{11} = 
0.56$ (95\% CI: 0.38, 0.80) and $\hat{\theta}_{10} = 0.003$ (95\% 
CI: 0.000, 0.008) are substantially lower than those for 
\textit{Dactylorhiza fuchsii}, which is ecologically plausible: 
\textit{Carex limosa} is a less charismatic species occupying remote 
upland bog habitats, resulting in lower recorder interest and 
reporting rates relative to the more accessible habitats favoured by 
\textit{Dactylorhiza fuchsii} --- all of which are absorbed by 
$\hat{\theta}_{11}$. The near-zero $\hat{\theta}_{10}$ confirms that 
false positive records are negligible for this species, consistent 
with its distinctive morphology making misidentification unlikely.
\begin{table}[H]
\centering
\caption{Case study 2: Results for DO-M applied to two vascular 
plant species in the United Kingdom. Species~1 is \textit{Dactylorhiza 
fuchsii} (common spotted orchid) and Species~2 is \textit{Carex 
limosa} (bog-sedge). The mean, standard deviation (SD), and 95\% 
posterior credible intervals (CI) are reported for the intercept, 
coefficients of environmental conditions, and observation process 
parameters $\theta_{11}$ and $\theta_{10}$.}
\label{tab:uk_po1}

\centering
\begin{tabular}[t]{lrrlrrl}
\toprule
\multicolumn{1}{c}{ } & \multicolumn{3}{c}{\textit{Species 1 (Orchis fuchsii)}} & \multicolumn{3}{c}{\textit{Species 2 (Carex limosa)}} \\
\cmidrule(l{3pt}r{3pt}){2-4} \cmidrule(l{3pt}r{3pt}){5-7}
\multicolumn{1}{c}{Parameters} & \multicolumn{1}{c}{Mean} & \multicolumn{1}{c}{SD} & \multicolumn{1}{c}{95\% CI} & \multicolumn{1}{c}{Mean} & \multicolumn{1}{c}{SD} & \multicolumn{1}{c}{95\% CI}\\
\midrule
Intercept & 14.728 & 1.494 & (11.771, 17.678) & -6.902 & 1.663 & (-10.157, -3.671)\\
Rain & 0.193 & 0.109 & (-0.019, 0.409) & 0.352 & 0.214 & (-0.010, 0.831)\\
Temperature & 0.537 & 0.156 & (0.238, 0.848) & 1.081 & 0.269 & (0.589, 1.642)\\
Dry Nitrogen & 0.936 & 0.168 & (0.612, 1.270) & -0.456 & 0.265 & (-0.988, 0.050)\\
Wet Nitrogen & 0.227 & 0.122 & (-0.007, 0.473) & -0.112 & 0.215 & (-0.540, 0.309)\\
GenomeSize & 0.121 & 0.121 & (-0.114, 0.355) & -0.521 & 0.226 & (-0.967, -0.084)\\
Bog & -0.230 & 0.363 & (-0.937, 0.474) & 1.498 & 0.440 & (0.672, 2.363)\\
ImprovedGrassland & 0.985 & 0.271 & (0.466, 1.541) & -0.792 & 0.358 & (-1.494, -0.106)\\
AcidNeutralGrassland & 0.814 & 0.318 & (0.211, 1.437) & 0.858 & 0.430 & (0.028, 1.703)\\
CalcareousGrassland & 1.007 & 0.713 & (-0.296, 2.489) & -0.384 & 0.868 & (-2.133, 1.249)\\
Rocky & -1.541 & 0.662 & (-2.915, -0.288) & -1.101 & 0.816 & (-2.698, 0.471)\\
ArableHorticulture & 0.745 & 0.285 & (0.185, 1.318) & -2.697 & 0.554 & (-3.847, -1.667)\\
UrbanSuburban & 1.062 & 0.504 & (0.143, 2.115) & -1.192 & 0.673 & (-2.549, 0.077)\\
SedimentCoastal & 0.202 & 0.389 & (-0.542, 0.984) & -1.397 & 0.642 & (-2.713, -0.215)\\
Heathland & 0.069 & 0.287 & (-0.486, 0.645) & 1.397 & 0.394 & (0.649, 2.205)\\
ConiferousWoodland & 0.996 & 0.347 & (0.326, 1.691) & 1.021 & 0.457 & (0.146, 1.934)\\
BroadleavedWoodland & 0.943 & 0.747 & (-0.461, 2.462) & 0.077 & 0.824 & (-1.619, 1.650)\\
Latitude & -0.265 & 0.026 & (-0.317, -0.212) & 0.091 & 0.031 & (0.030, 0.152)\\
$\theta_{11}$ & 0.931 & 0.018 & (0.896, 0.967) & 0.555 & 0.111 & (0.379, 0.801)\\
$\theta_{10}$ & 0.027 & 0.018 & (0.003, 0.071) & 0.003 & 0.002 & (0.000, 0.008)\\
\bottomrule
\end{tabular}

\end{table}

\section{Discussion}\label{sec:discussion}

The statistical literature has long treated occurrence models, species 
distribution models, and detection-only methods as fundamentally 
distinct approaches, developed in parallel to address what appeared 
to be separate problems. In this paper, we have argued that these are 
all special cases of a single hierarchical observation process, and 
introduced a unified terminology --- detection/non-detection data with 
$T$ visits (\text{DN-}$T$) and detection-only data (\text{DO}) --- 
that makes these connections explicit. Within this framework, we have 
proposed a novel hierarchical model for \text{DO} data, DO-M, that 
for the first time accounts for false positive errors in \text{DO} 
data, and accounts for false negative errors without requiring 
separate detection covariates --- which, whilst they can be 
incorporated following \citet{griffin2020modelling}, are rarely 
available for \text{DO} data in practice. The key insight is that 
the observation process underlying \text{DO} data is structurally 
analogous to that of \text{DN-1}, with the critical difference that 
the visit probability $q$ is unknown and non-detections are not 
recorded. This means that $p_{11}$ and $p_{10}$ are confounded with 
$q$ in \text{DO} data, so that only the products $\theta_{11} = 
qp_{11}$ and $\theta_{10} = qp_{10}$ are identifiable from the data 
alone. Recognising this confounding, and resolving it through 
ecologically justifiable informative prior distributions on $p_{11}$ 
and $p_{10}$, is what allows consistent estimation of the 
coefficients of environmental conditions on occurrence probability 
from \text{DO} data.

The unified terminological framework introduced in this paper has 
value beyond its role as motivation for DO-M. By making explicit that 
these approaches are all modelling the same underlying latent process 
--- the occurrence state $Z_i$ and its dependence on environmental 
conditions --- the framework clarifies what each data type can and 
cannot tell us. \text{DN-}$T$ data, with their known visit structure 
and repeated observations, provide the richest information about the 
observation process and allow $p_{11t}$ and $p_{10t}$ to be estimated 
without informative prior distributions. \text{DN-1} data provide 
less information about the observation process but retain the crucial 
advantage that non-detections are recorded, allowing the distinction 
between an unvisited site and a visited but unoccupied site to be 
made. \text{DO} data, being the least structured, carry the most 
uncertainty: a missing record is genuinely ambiguous, and $q$ is 
unknown. Positioning these data types within a single framework makes 
clear that the choice of modelling approach should be driven by the 
data that are available and the assumptions that can be justified, 
rather than by the historical accident of separate methodological 
traditions developing in parallel. We hope that the 
\text{DN-}$T$/\text{DN-1}/\text{DO} terminology will prove useful 
to practitioners and researchers navigating this literature.

The formal connection between \text{DO} and \text{DN-1} data 
established in this paper --- that the DO-M likelihood has the same 
form as the DN-1 likelihood, with $\theta_{11}$ and $\theta_{10}$ in 
place of $p_{11}$ and $p_{10}$ --- is a new insight that reframes 
how \text{DO} data should be modelled. Previous approaches to 
\text{DO} data, including Maxent \citep{phillips2004maximum}, the 
IPPP \citep{warton2010poisson}, and logistic regression with 
background samples \citep{ward2009presence}, all implicitly set $q = 
1$ or ignore the visit process entirely, thereby treating detections 
as true presences and ignoring false positive errors. Our simulation 
study demonstrates the consequences: these approaches yield severely 
biased estimates of the coefficients of environmental conditions, 
with credible interval coverage collapsing to zero across all 
simulation scenarios considered. In contrast, DO-M achieves 
estimation performance comparable to DN-1-M despite \text{DO} data 
being substantially less informative, confirming that the additional 
uncertainty inherent in \text{DO} data --- unknown $q$, unrecorded 
non-detections --- is properly propagated through the model.

A widely used approach for applying occurrence models to \text{DO} 
data is benchmarking \citep{Isaac2014, Altwegg2019}, in which 
detection records of other species at the same sites are used to 
infer that a site was visited even when the focal species was not 
recorded, thereby constructing non-detections for use in a standard 
occurrence model. Whilst benchmarking is practically appealing, it 
rests on a critical assumption within our framework: that $q$ is the 
same across species, so that a site visited for one species is 
equally likely to have been properly surveyed for the focal species. 
Given that $q$ absorbs reporting behaviour and recorder interest in 
addition to visit probability, this assumption is violated whenever 
species differ in charisma or recorder appeal --- precisely the 
conditions that characterise most multi-species citizen science 
datasets. Moreover, benchmarking does not propagate the uncertainty 
arising from unrecorded non-detections, and entirely ignores the 
visiting process in the manner formalised in our framework. 
\text{DO-M} avoids these assumptions by treating $q$ as an unknown 
composite quantity to be estimated, with uncertainty properly 
propagated through the posterior distribution.

A notable finding of the simulation study is that the well-known 
divergence of the intercept term as background sample size grows 
\citep{owen2007infinitely, renner2013equivalence} disappears under 
DO-M. This divergence, which \cite{renner2013equivalence} attribute 
to model misspecification, is indeed a symptom of misspecification: 
when observation error is ignored, the likelihood cannot distinguish 
between a site with no detection because it was not visited and a 
site with no detection because the species is absent, leading to an 
increasingly poorly specified model as background sample size grows. 
Under DO-M, which correctly accounts for the observation process, 
both the intercept and the coefficients converge to their true values 
as background sample size grows, with coefficients converging faster 
than the intercept. The practical implication is that practitioners 
using DO-M can use large background samples --- as recommended by 
\cite{barbet2012selecting} for regression and machine learning 
approaches --- without inducing the pathological intercept behaviour 
observed under naive approaches.

The two case studies illustrate the practical value of DO-M. In the 
New Zealand case study, where both \text{DN-1} and \text{DO} data 
are available for the same species, the close agreement between the 
results obtained from DN-1-M and DO-M applied to species 
\textit{nz42} is reassuring: DO-M recovers virtually the same 
important environmental drivers as DN-1-M, with only minor 
differences in the credible intervals. This suggests that the 
information lost by having only \text{DO} data rather than 
\text{DN-1} data is modest.

In the United Kingdom case study, the model yields ecologically 
interpretable results for both species. For \textit{Dactylorhiza fuchsii}, 
the identified drivers --- temperature, dry nitrogen deposition, acid 
and neutral grassland, improved grassland, rocky habitat, arable and 
horticulture, urban and suburban areas, coniferous woodland, and 
latitude --- are consistent with the known ecology of this widespread 
orchid, which tolerates a broad range of managed and semi-natural 
habitats across England, Wales, and southern Scotland.

For \textit{Carex limosa}, the identified drivers --- temperature, 
genome size, bog, improved grassland, acid and neutral grassland, 
arable and horticulture, sediment coastal, heathland, coniferous 
woodland, and latitude --- strongly reflect the species' 
specialisation to waterlogged, nutrient-poor habitats and its 
limited ability to persist in agriculturally improved or disturbed 
landscapes.

The estimated observation process parameters provide an additional 
layer of ecological insight. 
The substantially lower 
$\hat{\theta}_{11}$ for \textit{Carex limosa} relative to 
\textit{Dactylorhiza fuchsii} is ecologically plausible, reflecting lower 
recorder interest and reporting rates for this less charismatic 
species occupying remote upland bog habitats. The small estimated 
false positive rate for \textit{Dactylorhiza fuchsii} reflects the known 
difficulty of distinguishing this species from other members of the 
genus \textit{Dactylorhiza}, such as \textit{Dactylorhiza maculata}, 
which shares close morphological resemblance and is also widely 
distributed across the United Kingdom. The near-zero 
$\hat{\theta}_{10}$ for \textit{Carex limosa} confirms that false 
positive records are negligible for this species, consistent with 
its distinctive morphology making misidentification unlikely.

DO-M has several limitations that should be acknowledged. First, the 
model relies on informative prior distributions for $p_{11}$ and 
$p_{10}$ to achieve identifiability: if these prior distributions 
are badly misspecified, the resulting estimates of the coefficients 
of environmental conditions may be unreliable. In practice, 
empirical studies of observer accuracy 
\citep{groom2017characterisation} can guide prior distribution 
specification, but such studies are not available for all taxa. In 
such cases, the Sheffield Elicitation Framework (SHELF) 
\citep{o2006uncertain} provides a structured alternative for 
eliciting prior information about $p_{11}$ and $p_{10}$ from domain 
experts. Second, the model assumes that visits occur independently 
of the latent occurrence state $Z_i$. This assumption may be 
violated when recorders preferentially visit sites where they expect 
to find the species of interest --- a form of preferential sampling 
that is well-documented in citizen science data 
\citep{chakraborty2011point}. Relaxing this assumption is an 
important direction for future work. Third, DO-M in its current form 
assumes that the coefficients of environmental conditions are 
constant across space. In practice, species--environment 
relationships often vary spatially due to interactions with 
unmeasured biotic and abiotic processes operating at different 
scales \citep{Doser2024}. In our case studies, we partially address 
this by latitude-adjusting bio-climatic covariates that exhibit 
strong latitudinal trends. However, this does not fully account for 
spatial non-stationarity in covariate effects, and a natural 
extension of DO-M would incorporate spatially varying coefficients, 
for example through a Gaussian process specification 
\citep{Doser2024}, though this represents a technically challenging 
direction given the computational demands of Gaussian process models 
at the spatial scales typical of \text{DO} datasets. Fourth, the 
model does not account for spatial autocorrelation in occurrence --- 
the tendency for nearby sites to have similar occurrence states 
regardless of measured environmental conditions. Ignoring spatial 
autocorrelation can lead to underestimated uncertainty in coefficient 
estimates \citep{ketwaroo2024models}, and the model can be extended 
to include spatial random effects, for example via a conditional 
autoregressive (CAR) model, which is straightforward to implement in 
Stan \citep{cmdstan}. Fifth, DO-M is more computationally demanding 
than machine learning approaches such as Maxent, BRT, and RF: whilst 
the latter can be fitted in seconds, DO-M requires Hamiltonian Monte 
Carlo sampling which may take considerably longer for large datasets. 
However, as our simulation study demonstrates, the speed advantage 
of machine learning approaches comes at a considerable cost: 
coefficient estimates are severely biased and credible interval 
coverage collapses to zero, making these approaches unreliable for 
inference on the drivers of species distributions. For applications 
where reliable inference is the primary objective --- which we argue 
should be the norm in ecology --- DO-M provides the only principled 
and statistically coherent alternative among the methods considered 
here.

Several directions for future work present themselves. The unified 
framework introduced in this paper provides a natural basis for 
integrating \text{DN-}$T$ and \text{DO} data within a single model, 
which would allow the complementary strengths of each data type to 
be exploited --- the richer observation process information in 
\text{DN-}$T$ data and the greater spatial coverage of \text{DO} 
data. Integration of multiple data types within a unified occurrence 
modelling framework is an active area of research 
\citep{isaac2020, miller2019joint} and the connections established 
in this paper provide a principled foundation for such extensions. 
The model could also be extended to accommodate temporal dynamics 
in occurrence, relaxing the closed-population assumption and 
allowing $Z_i$ to change over time, which would be particularly 
valuable for monitoring applications where changes in species 
distributions over time are of primary interest. Finally, the 
framework could be extended to multiple species, sharing information 
across species on the observation process parameters $p_{11}$ and 
$p_{10}$ through hierarchical prior distributions, which would be 
particularly useful when species-specific empirical estimates of 
observer accuracy are not available.

In conclusion, this paper has established that detection-only 
methods, species distribution models, and traditional occurrence 
models are all special cases of a single hierarchical observation 
process, and has proposed the first principled model for \text{DO} 
data that accounts for both false positive and false negative errors. 
The proposed model provides reliable estimates of the coefficients 
of environmental conditions on occurrence probability from \text{DO} 
data, with well-calibrated uncertainty quantification, and opens the 
door to a more unified and statistically coherent approach to species 
occurrence modelling.
\newpage
\appendix
\appendix
\renewcommand{\thesection}{\arabic{section}}

\section{Supplementary Material}
We provide technical details of occupancy models for detection/non-detection data and their inevitable identification conditions. We also provide technical details of hitherto detection-only methods, such as the Maximum entropy method, logistic regression model, and the inhomogeneous Poisson point process model. In addition, we provide additional results of our simulation studies, case studies, and details of the data used.   

\subsection{Identifiability of \text{DN-}$T$ and \text{DN-1} 
            models}\label{supp:dnt}

\subsubsection{Detection/non-detection data with $T > 1$ visits}\label{supp:dn-t}

The log-likelihood for \text{DN-}$T$ data is
\begin{equation}
\ell(p_{11t},\,p_{10t},\,\boldsymbol{\beta} \mid \mathbf{Y})
= \sum_{i=1}^{n} \sum_{t:\,V_{it}=1}
\log \Big[
\big(p_{11t}\psi_i + p_{10t}(1-\psi_i)\big)^{Y_{it}}
\big(1 - p_{11t}\psi_i - p_{10t}(1-\psi_i)\big)^{1-Y_{it}}
\Big],
\label{eq:supp_dnt}
\end{equation}
where the sum over $t$ is restricted to occasions on which site $s_i$ 
was visited. This is described extensively in 
\cite{mackenzie2017occupancy}. The special case $p_{10t} = 0$ for all 
$t$, presented by \cite{mackenzie2002estimating}, constrains the model 
so that all detections are true presences and only false negative errors 
are possible. \cite{royle2006generalized} relaxed this assumption to 
allow $p_{10t} > 0$, permitting both false positive and false negative 
errors simultaneously, and also proposed a generalisation for more than 
two categories, i.e., $Z_i \in \{0, 1, 2, \ldots\}$ and 
$Y_{it} \in \{0, 1, 2, \ldots\}$.

The likelihood exhibits a symmetry: the parameter combinations 
$(p_{11t}, p_{00t}, -\boldsymbol{\beta})$ and 
$(p_{10t}, p_{01t}, \boldsymbol{\beta})$ yield identical likelihood 
values, i.e.,
\begin{equation}
\mathcal{L}(p_{11t}, p_{00t}, -\boldsymbol{\beta}) 
= \mathcal{L}(p_{10t}, p_{01t}, \boldsymbol{\beta}),
\end{equation}
as noted by \cite{royle2006generalized}, \cite{de2021need}, 
\cite{guillera2017dealing}, and \cite{griffin2020modelling}. This 
symmetry leads to a bimodal posterior distribution and is resolved by 
imposing the constraint $p_{11t} > p_{10t}$ for all $t$. The reverse 
constraint $p_{11t} < p_{10t}$ also breaks the symmetry but leads to a 
reverse interpretation of the coefficients $\boldsymbol{\beta}$, which 
is ecologically undesirable. The multimodality issue becomes more complex 
for the unconstrained model ($p_{10t} > 0$) when 
$Z_i \in \{0, 1, 2, \ldots\}$ \citep{royle2006generalized}; however, 
in this paper we limit our analysis to $Z_i \in \{0, 1\}$, where model 
identification is less complicated.

\subsubsection{Detection/non-detection data with a single visit} 
\label{supp:dn1}

The log-likelihood for \text{DN-1} data is
\begin{equation}
\ell(p_{11},\,p_{10},\,\boldsymbol{\beta} \mid \mathbf{Y})
= \sum_{i=1}^{n} 
\log \Big[ 
\big(p_{11}\psi_i + p_{10}(1-\psi_i)\big)^{Y_i} 
\big(1 - p_{11}\psi_i - p_{10}(1-\psi_i)\big)^{1-Y_i} 
\Big].
\label{eq:supp_dn1}
\end{equation}
The special case $p_{10} = 0$, as in \cite{lele2012dealing}, reduces 
this to
\begin{equation}
\ell(p_{11},\,\boldsymbol{\beta} \mid \mathbf{Y})
= \sum_{i=1}^{n} 
\log \Big[ 
(p_{11}\psi_i)^{Y_i} 
\big(1 - p_{11}\psi_i\big)^{1-Y_i} 
\Big],
\label{eq:supp_dn1_p10zero}
\end{equation}
under which modelling occupancy and detection with separate covariate 
sets $\mathbf{x}_i$ and $\mathbf{v}_i$ becomes necessary. 
\cite{solymos2012conditional} and \cite{lele2006weighted} establish 
identifiability of this model under the logit link when $\mathbf{x}_i$ 
and $\mathbf{v}_i$ are sufficiently different. However, 
\cite{knape2015estimates} showed that there exist alternative link 
functions that generate data with the same distribution as $Y_i$, for 
example when the inverse logit is scaled by a constant. Therefore, 
occupancy probability is not identifiable in general; only the relative 
rather than absolute occupancy probability can be reliably estimated. 
Reliable estimates of the slope parameters can be obtained when 
$\mathbf{x}_i$ and $\mathbf{v}_i$ are completely disjoint 
\citep{knape2015estimates}. When the log-link function is used, the 
intercepts $\beta_0$ and $\delta_0$ for occupancy and detection are 
confounded and cannot be estimated separately \citep{knape2015estimates}, 
and there is additional confounding between coefficients of covariates 
present in both $\mathbf{x}_i$ and $\mathbf{v}_i$.
\subsection{Detection-only data}\label{supp:do}
\subsubsection{Technical details: Maxent}\label{supp:maxent}

The Maxent model \citep{phillips2004maximum, phillips2008modeling} 
estimates the target distribution $\pi_1(s)$ by minimising the 
Kullback-Leibler divergence
\begin{equation}
H(\pi_1) = \int_{\mathcal{A}} \pi_1(s)\log\!\left(
\frac{\pi_1(s)}{\pi_{\mathcal{A}}(s)}\right)ds
\end{equation}
subject to the constraints
\begin{equation}
\int_{\mathcal{A}} \pi_1(s)\,ds = 1, \qquad
\int_{\mathcal{A}} \pi_1(s)\,\mathbf{x}\,ds
= \frac{1}{n_1}\sum_{i=1}^{n_1}\mathbf{x}_i,
\label{eq:supp_maxent_constraints}
\end{equation}
that is, the estimated density $\pi_1(s)$ must integrate to one and 
its implied mean of $\mathbf{x}$ must match the empirical mean at 
detection sites. The solution to this minimisation problem is a Gibbs 
distribution \citep{phillips2006maximum}, which is an exponential 
family model
\begin{equation}
\pi_1(s) = \pi_{\mathcal{A}}(s)\,
e^{\beta_0 + \boldsymbol{\beta}^\top\mathbf{x}},
\end{equation}
where $\beta_0$ is a normalising constant ensuring 
$\int_{\mathcal{A}}\pi_1(s)\,ds = 1$ \citep{elith2011statistical}. 
The Maxent solution for $\pi_1(s)$ is therefore
\begin{equation}
\pi_1(s) = \frac{\pi_{\mathcal{A}}(s)\,e^{\boldsymbol{\beta}^\top\mathbf{x}}}
{\int_{\mathcal{A}}\pi_{\mathcal{A}}(u)\,
e^{\boldsymbol{\beta}^\top\mathbf{x}(u)}\,du}.
\end{equation}
With a large background sample $n_0$, $\pi_{\mathcal{A}}$ is replaced 
with a uniform distribution $1/n_0$ and the approximation for 
$\pi_1(s)$ becomes
\begin{equation}
\pi_1(s) \approx \frac{e^{\boldsymbol{\beta}^\top\mathbf{x}}}
{\sum_{i=n_1+1}^{n_0}e^{\boldsymbol{\beta}^\top\mathbf{x}_i}}.
\end{equation}
Since $\pi_1(s) = \mathbb{P}(\mathbf{x} \mid Y_i = 1)$ and 
$\pi_{\mathcal{A}}(s) = \mathbb{P}(\mathbf{x})$, the conditional 
probability $\mathbb{P}(Y_i = 1 \mid \mathbf{x})$ is obtained by 
invoking Bayes' rule \citep{ward2007statistics}:
\begin{equation}
\mathbb{P}(Y_i = 1 \mid \mathbf{x}) 
= \frac{\pi_1(s)}{\pi_{\mathcal{A}}(s)} \cdot \mathbb{P}(Y_i = 1).
\label{eq:supp_bayes}
\end{equation}
Maxent computes the ratio $\pi_1(s)/\pi_{\mathcal{A}}(s)$, known as 
the raw output, but $\tau = \mathbb{P}(Y_i = 1)$ --- the proportion 
of occupied area (PAO) --- is not known from \text{DO} data alone and 
is required to recover $\mathbb{P}(Y_i = 1 \mid \mathbf{x})$. To 
circumvent this, the log of the raw output is treated as logit scores, 
and the average of these scores at detection sites is assumed to be 
$0.5$ by default. See \cite{phillips2004maximum}, 
\cite{phillips2006maximum}, and \cite{elith2011statistical} for 
further details.

\subsubsection{Technical details: logistic regression with background 
            samples}\label{supp:logreg}

The case-control adjusted logistic regression model 
\citep{ward2009presence, dorazio2012predicting} models 
$\mathbb{P}(Y_i = 1 \mid \mathbf{x})$ using \text{DO} data combined 
with background samples. Using Bayes' rule,
\begin{equation}
\text{logit}\big(\mathbb{P}(Y_i = 1 \mid \mathbf{x})\big)
= \log\!\left(\frac{\pi_1(s)}{\pi_0(s)}\right) 
+ \log\!\left(\frac{\tau}{1-\tau}\right),
\label{eq:supp_logit}
\end{equation}
where $\tau = \mathbb{P}(Y_i = 1)$ and $\pi_0(s) = 
\mathbb{P}(\mathbf{x} \mid Y_i = 0)$. \cite{ward2007statistics} 
demonstrated that $\tau$ and $\pi_0(s)$ are not uniquely defined 
without further assumptions on the structure of 
$\eta_i(\mathbf{x})$, and proposed estimating the underlying latent 
detections/non-detections from background data via the 
Expectation-Maximisation (EM) algorithm of \cite{dempster1977maximum}. 
The algorithm proceeds as follows:
\begin{enumerate}
\item Fix the \text{DO} data and assign a prior value for $\tau$ to 
background sites.
\item M-step: fit the case-control logistic regression model and 
apply the case-control adjustment 
$-\log\!\left((n_1 + \tau n_0)/((1-\tau)n_0)\right) 
- \log(\tau/(1-\tau))$ to the intercept to obtain 
$\eta_i(\mathbf{x})$.
\item E-step: compute the expected latent detections/non-detections 
in the background data.
\item Repeat until convergence.
\end{enumerate}
Where $n_1 + \tau n_0$ and $(1-\tau)n_0$ are the expected number of 
detections and non-detections, respectively. The shrinkage of 
coefficient estimates relative to the naive logistic regression model 
is reduced since they are re-estimated at each iteration. However, 
the model assumes perfect detections for the \text{DO} data, which 
still has consequences for the coefficient estimates. If $\tau$ is 
unknown, the naive logistic regression model is the best available 
alternative \citep{ward2007statistics, ward2009presence}. The log-likelihood of the naive logistic regression model is
\begin{equation}
\ell(\beta_0, \boldsymbol{\beta})
= \sum_{i=1}^{n_1}
\left\{(\beta_0 + \boldsymbol{\beta}^\top\mathbf{x}) 
- \log\!\left(1 + e^{\beta_0 + \boldsymbol{\beta}^\top\mathbf{x}}
\right)\right\}
- \sum_{i=1}^{n_0}
\log\!\left(1 + e^{\beta_0 + \boldsymbol{\beta}^\top\mathbf{x}}
\right).
\label{eq:supp_nllr}
\end{equation}
\cite{owen2007infinitely} showed that the last term in 
Equation~\ref{eq:supp_nllr} can be replaced by
\begin{equation}
n_0\int_{\mathcal{A}}
\log\!\left(1 + e^{\beta_0 + \boldsymbol{\beta}^\top\mathbf{x}}
\right)\pi_{\mathcal{A}}(s)\,ds
\end{equation}
to obtain the log-likelihood for the infinitely imbalanced logistic 
regression (IILR), which is well approximated for large $n_0$. This 
demonstrates that whilst the slope parameters converge to their true 
values as $n_0 \to \infty$, the intercept diverges as 
$\beta_0 \to -\infty$. The existence of the MLE for the logistic 
regression model requires some overlap between the values of 
$\mathbf{x}$ for $Y_1$ and $Y_0$; \cite{silvapulle1981existence} 
showed that some degree of overlap is necessary, and 
\cite{owen2007infinitely} proposed an overlap condition for the case 
$n_0 \to \infty$ with finite $n_1$.

\subsubsection{Technical details: inhomogeneous Poisson point process 
            model}\label{supp:ippp}

\cite{warton2010poisson} proposed modelling \text{DO} data as 
realisations of an inhomogeneous Poisson point process (IPPP) with 
log-linear intensity function
\begin{equation}
\log(\lambda_i) = \beta_0 + \boldsymbol{\beta}^\top\mathbf{x},
\label{eq:supp_ippp_intensity}
\end{equation}
with log-likelihood
\begin{equation}
\ell(\beta_0, \boldsymbol{\beta})
= \sum_{i=1}^{n_1}(\beta_0 + \boldsymbol{\beta}^\top\mathbf{x})
- \int_{\mathcal{A}}\lambda(s)\,ds - \log(n_1!).
\label{eq:supp_ippp_lik}
\end{equation}
The integral is evaluated using quadrature points 
\citep{warton2010poisson}. \cite{fithian2012finite} showed that it 
can be approximated numerically using uniform background samples 
$|\mathcal{A}|/n_0$, giving
\begin{equation}
\ell(\beta_0, \boldsymbol{\beta})
= \sum_{i=1}^{n_1}(\beta_0 + \boldsymbol{\beta}^\top\mathbf{x})
- \frac{|\mathcal{A}|}{n_0}\sum_{i=1}^{n_0}\lambda_i - \log(n_1!).
\label{eq:supp_ippp_lik_approx}
\end{equation}
\cite{moreira2022analysis} proposed an extension of the IPPP for 
the occurrence-detection process, specifying two latent processes for 
the unobserved detections that lead to an analytically exact 
likelihood function, avoiding numerical approximation of the integral. 
Whilst \cite{moreira2022analysis} demonstrate that identifiability 
is achievable in this setting, they acknowledge that covariates for 
the detection process are not readily available for \text{DO} data 
in practice.

\text{DO} data reflect only the intensity of detections, not of true 
occurrences: the intensity of occurrence $\tilde{\lambda}(s)$ is 
thinned by the probability of detection $h(s)$, so that 
$\lambda = \tilde{\lambda}h$ \citep{chakraborty2011point}. There is 
confounding between $\tilde{\lambda}$ and $h$, so that 
$\tilde{\lambda}$ is not identifiable from \text{DO} data alone. 
Suppose log-linear models are specified for $\tilde{\lambda}$ and $h$ 
with independent covariate sets $\mathbf{x}$ and $\mathbf{v}$ and 
respective intercept and slope parameters. The intensity of detections 
is then also log-linear with combined covariates 
$\tilde{\mathbf{x}} = (\mathbf{x}, \mathbf{v})$, combined slopes 
$\tilde{\boldsymbol{\beta}} = (\boldsymbol{\beta}, \boldsymbol{\delta})$, 
and combined intercept $\tilde{\beta}_0 = \beta_0 + \delta_0$. 
\cite{fithian2012finite} noted that the slope parameters of $\lambda$ 
are identifiable, hence the slope parameters of $\tilde{\lambda}$ are 
also identifiable. However, $\beta_0$ cannot be estimated without 
knowing $\delta_0$, and the model is not identifiable if there is 
overlap between $\mathbf{x}$ and $\mathbf{v}$ 
\citep{fithian2012finite}. This identification issue is analogous to 
that of the \text{DN-1} model discussed in 
Section~\ref{supp:dn1}.

\subsection{Connections between Maxent, logistic regression, and 
            IPPP}\label{supp:connections}

\cite{renner2013equivalence} showed that Maxent 
\citep{phillips2004maximum} and the IPPP \citep{warton2010poisson} 
are equivalent, differing only in the intercept due to scale 
dependence in Maxent when the study area is divided into grid cells. 
Ignoring quadrature weights in the IPPP changes the intercept by 
$\log(|\mathcal{A}|/n_0)$, making it sensitive to resolution. 
\cite{fithian2012finite} further demonstrated that Maxent and the 
IPPP lead to the same probability density. \cite{fithian2012finite} 
also questioned the scientific relevance of the intercept term, 
showing that it is a normalising constant ensuring that $\lambda$ 
integrates to $n_1$; unless $n_1$ is of scientific interest, the 
intercept is not an object of inference.

The logistic regression model and IPPP are asymptotically equivalent 
when quadrature weights are ignored: see Theorems 3.1, 3.2, and 3.3 
of \cite{warton2010poisson}. With large background samples, the 
maximum likelihood estimates of slope parameters under the IPPP are 
achieved by the logistic regression model, but not the intercept. 
Under model misspecification, the logistic regression model requires 
larger background samples to achieve the same slope estimates as the 
IPPP. \cite{fithian2012finite} proposed the infinitely weighted 
logistic regression (IWLR) as a finite-sample equivalent of the IPPP. 
\cite{dorazio2012predicting} achieved similar results with the 
case-adjusted logistic regression.

Pseudo-absences in the logistic regression model, quadrature points 
in the IPPP, and background data in Maxent all serve the same 
purpose: to represent the distribution of environmental conditions 
across $\mathcal{A}$ and to approximate the integral over the study 
area in the likelihood. \cite{barbet2012selecting} recommended using 
at least 10,000 background samples for stable performance of 
regression and machine learning models, consistent with the default 
in Maxent \citep{elith2011statistical}. The IILR of 
\cite{owen2007infinitely} demonstrates that whilst holding $n_1$ 
fixed and increasing $n_0$, the intercept diverges as 
$n_0 \to \infty$ and the slope converges to its true value for the 
logistic regression model. Hence, the logistic regression model is 
not different from Maxent and IPPP in requiring large background 
samples for better approximation.

Although these models are known to provide consistent estimates of 
slope parameters under certain conditions, this consistency breaks 
down in the presence of detection errors 
\citep{dorazio2012predicting}. Specifically, consistent estimation 
requires disjoint covariate sets for the occurrence and detection 
processes (see Proposition 2 of \cite{dorazio2012predicting}), a 
condition that is rarely satisfiable in practice for \text{DO} data. 
The independence of $\mathbf{x}$ and $\mathbf{v}$ may even be 
impossible to establish \citep{fithian2012finite, moreira2022analysis}.

\subsection{Stan set up and sampling diagnostics}\label{supp:setup}

Our proposed model is written in the stan software using the 
\textbf{cmdstan} package in \textbf{R}; thus, the posterior samples are 
obtained using the Hamiltonian Monte Carlo (HMC) sampling algorithm 
implemented in Stan. For all simulations and case studies, we used four 
parallel chains, 2000 iterations for warmup and 2000 iterations for 
sampling per chain. As usual, there is a need for a proper 
initialisation of the sampler; generally, the observation parameters 
$p_{11}$, $p_{10}$, $\theta_{11}$, $\theta_{10}$ and $q$ should be set 
to some initial value to avoid these warnings ``Rejecting initial value" 
and ``Log probability evaluates to $\log(0)$" when using the Stan 
default initialisation. This warning occurs only in the first few iterations before sampling and does not appear during sampling. 

For both case studies, that is, New Zealand and the United Kingdom, the
continuous variables are normalised to have a mean of zero and a standard deviation of one. We use a uniform Beta prior for $q$, that is, $q \sim \text{Beta}(1, 1)$, $p_{11} \sim \mathrm{Beta}(5, 2)$, $p_{10} \sim \mathrm{Beta}(2, 20)$, $\beta_0 \sim N(0, 4)$, and standard normal distributions for the 
coefficients ($\beta_k \sim N(0, 1) : k = 1, 2, \dotsi, p$) of our environmental covariates. However, 
inference on the intercept ($\beta_0$) and effect sizes ($\beta_k : k = 1, 2, \dotsi, p$) is obtained using 
$\theta_{11} = qp_{11}$ and $\theta_{10} = qp_{10}$ derived by the change of variable theorem \citep{siegrist2024}. Alternatively, the Sheffield elicitation frame \cite{o2006uncertain} can be used to elicit prior information about species of interest from 
domain experts, especially, when empirical estimates of $p_{11}$ and 
$p_{10}$ are not available. 

All Stan diagnostics are okay, for both simulation and case studies.
The $\hat{R}$ (comparing between and within chain variations) is always
1, showing convergence of the four parallel chains, that is, all four 
chains end up in the same space. The effective sample sizes (ESS) \textit{ESS 
bulk} and \textit{ESS tail} are both greater than 2000 for all parameter 
estimates of the model, which is significantly more than 100\% of the number of chains and implies reliable estimates for posterior 
mean and credible intervals. See for more information on Stan diagnostics see: https://mc-stan.org/learn-stan/diagnostics-warnings.html.

\subsection{Priors and induced priors for observation parameters}\label{supp:induced-priors}
\begin{figure}[H]
    \centering
    \begin{subfigure}[t]{0.4\textwidth}
        \centering
        \includegraphics[width=\linewidth]
        {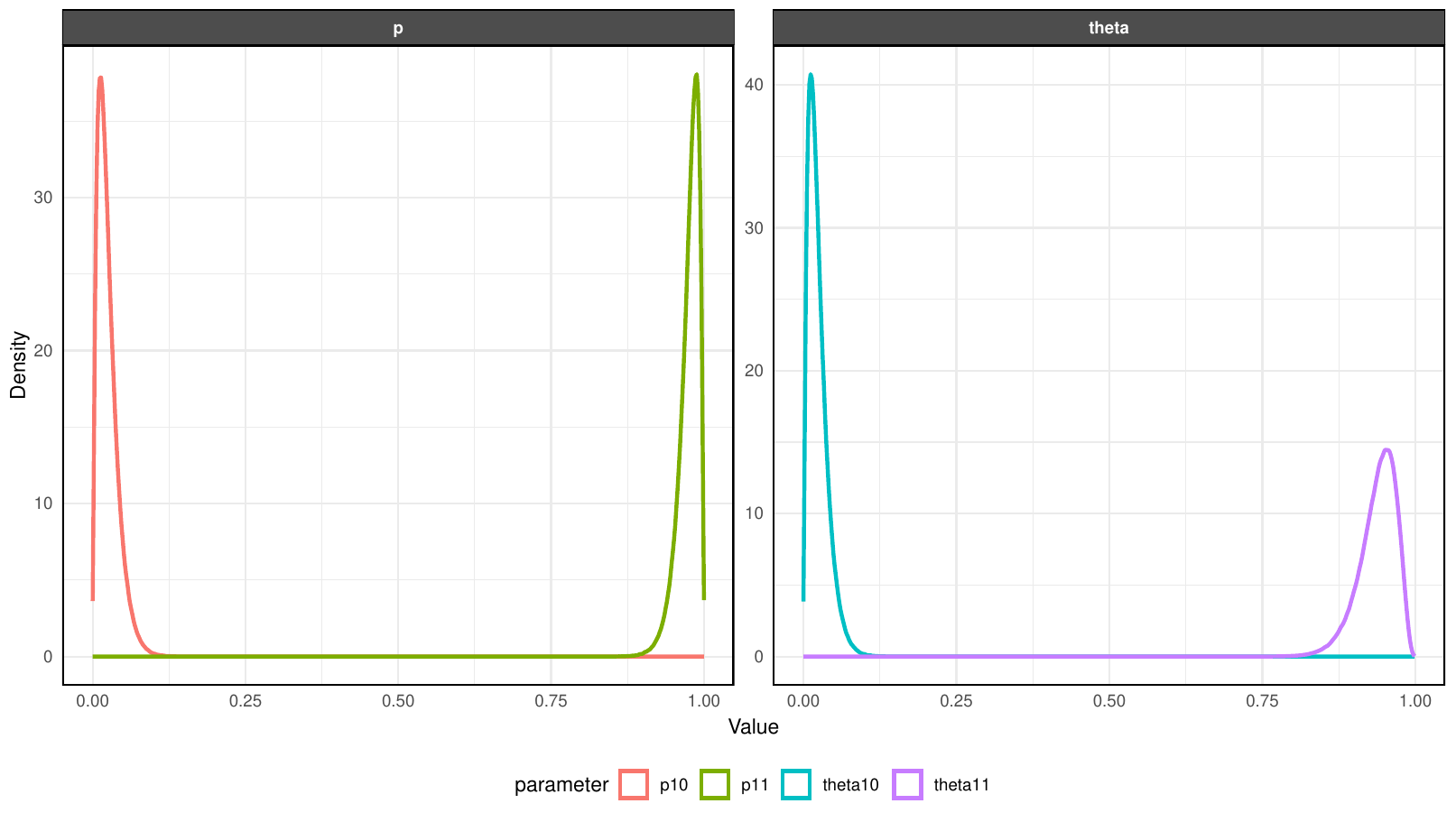}
        \caption{
        $p_{11}\sim \mathrm{Beta}(80,2)$,
        $p_{10}\sim \mathrm{Beta}(2,80)$, and $q \sim \mathrm{Beta}(50, 2)$.}
        \label{fig:info-p-q1}
    \end{subfigure}
    \hspace{0.02\textwidth}
     \begin{subfigure}[t]{0.4\textwidth}
        \centering
        \includegraphics[width=\linewidth]
        {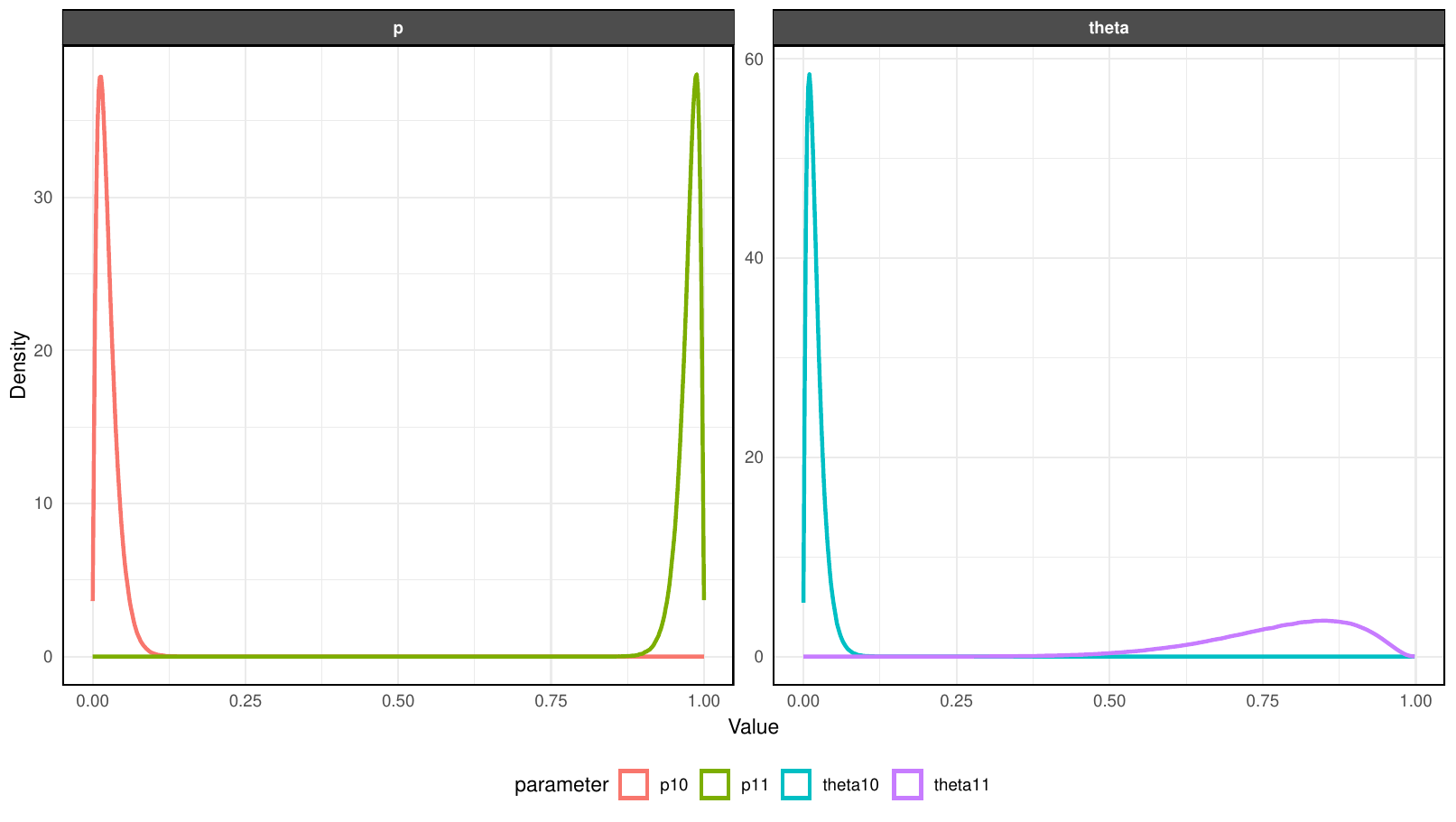}
        \caption{
        $p_{11}\sim \mathrm{Beta}(80,2)$,
        $p_{10}\sim \mathrm{Beta}(2,80)$, and $q \sim \mathrm(5, 2)$.}
        \label{fig:info-p-q2}
    \end{subfigure}
    \hspace{0.02\textwidth}
    \begin{subfigure}[t]{0.4\textwidth}
        \centering
        \includegraphics[width=\linewidth]
        {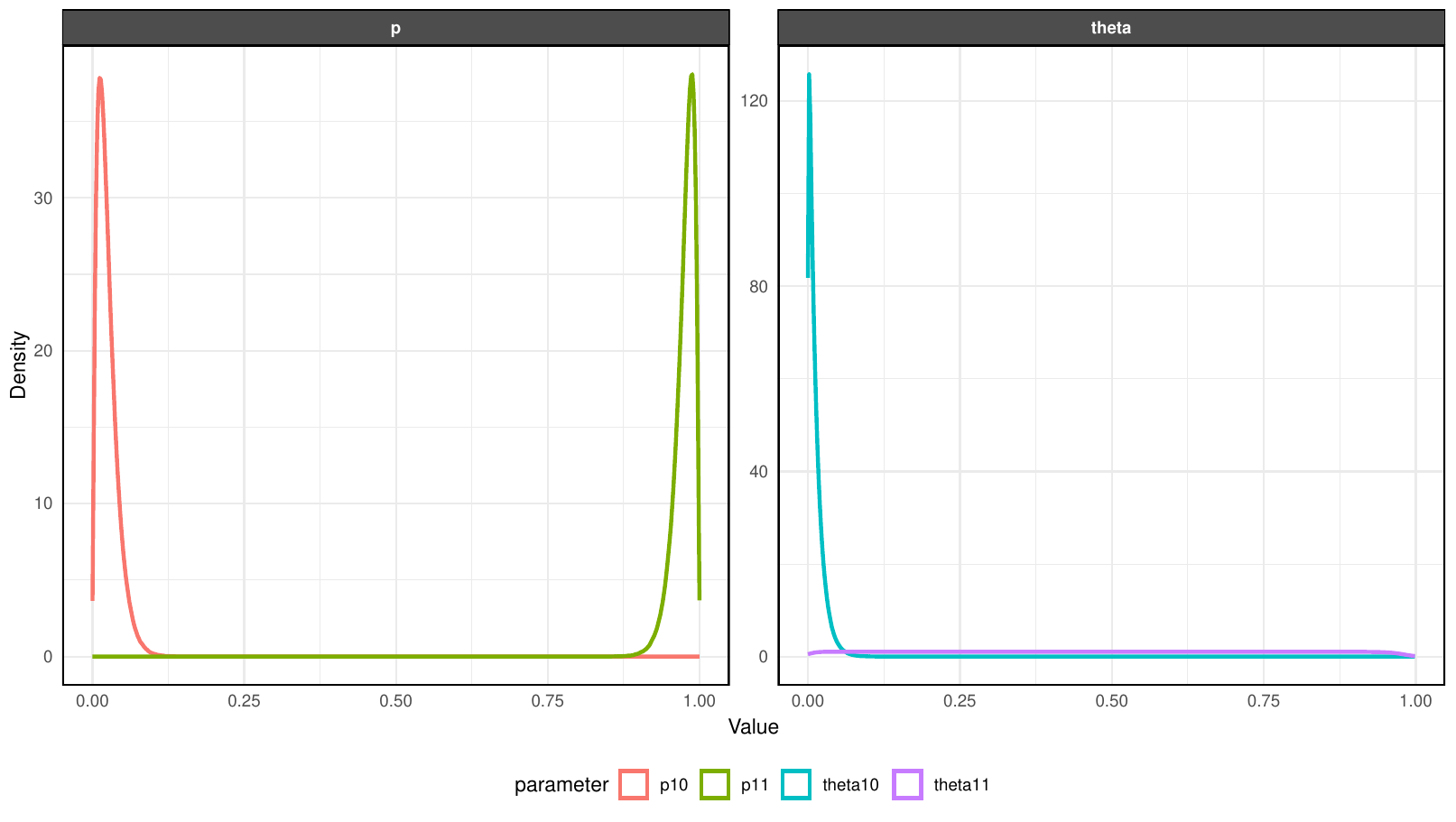}
        \caption{
        $p_{11}\sim \mathrm{Beta}(80,2)$,
        $p_{10}\sim \mathrm{Beta}(2,80)$, and $q \sim \mathrm{Beta}(1, 1)$.}
        \label{fig:informative}
    \end{subfigure}
    \hspace{0.02\textwidth}
    \begin{subfigure}[t]{0.4\textwidth}
        \centering
        \includegraphics[width=\linewidth]
        {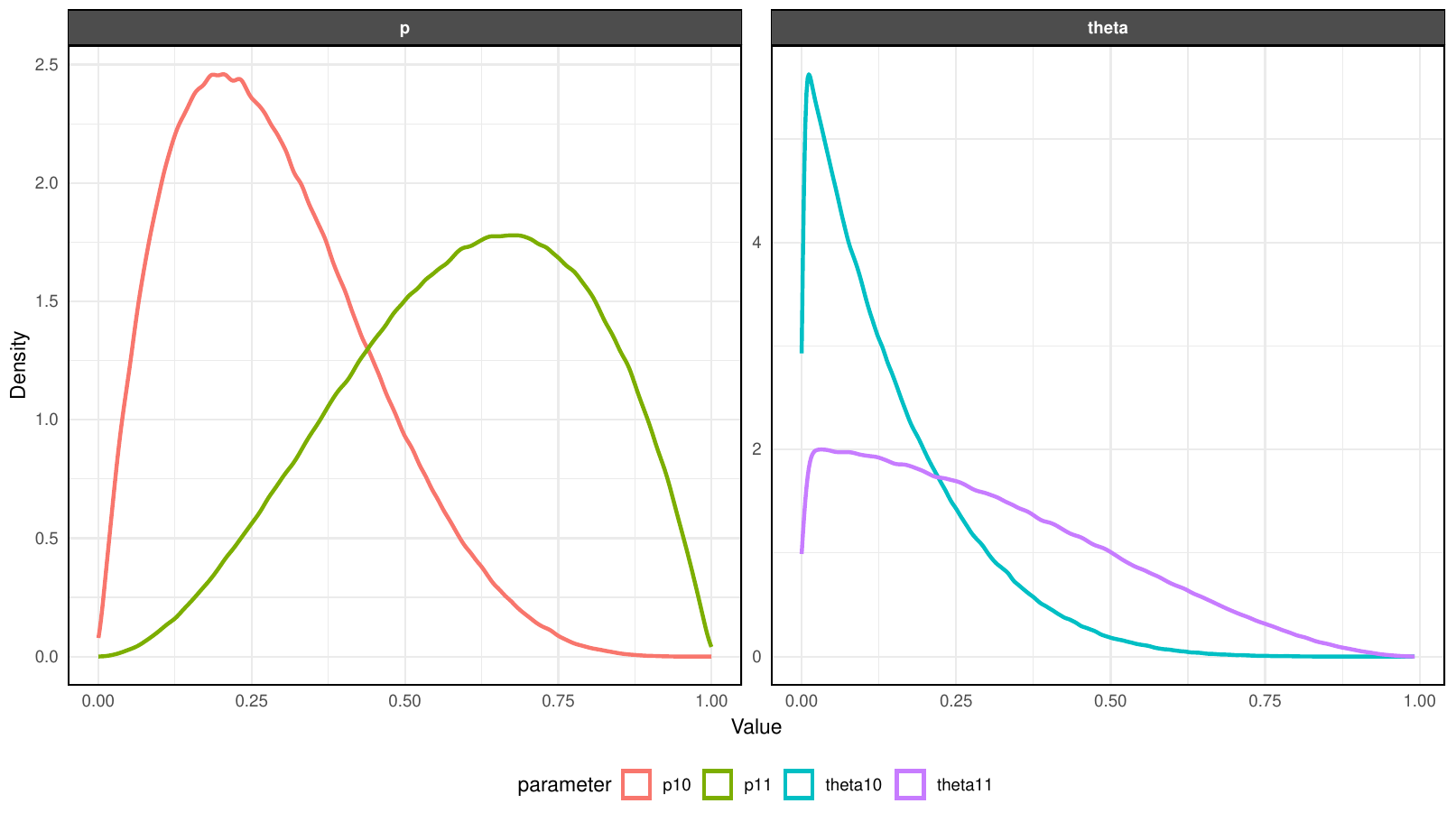}
        \caption{
        $p_{11}\sim \mathrm{Beta}(3,2)$,
        $p_{10}\sim \mathrm{Beta}(2,5)$, and $q \sim \mathrm{Beta}(1, 1)$.}
        \label{fig:noninformative}
    \end{subfigure}
    \hspace{0.02\textwidth}
    \begin{subfigure}[t]{0.4\textwidth}
        \centering
        \includegraphics[width=\linewidth]
        {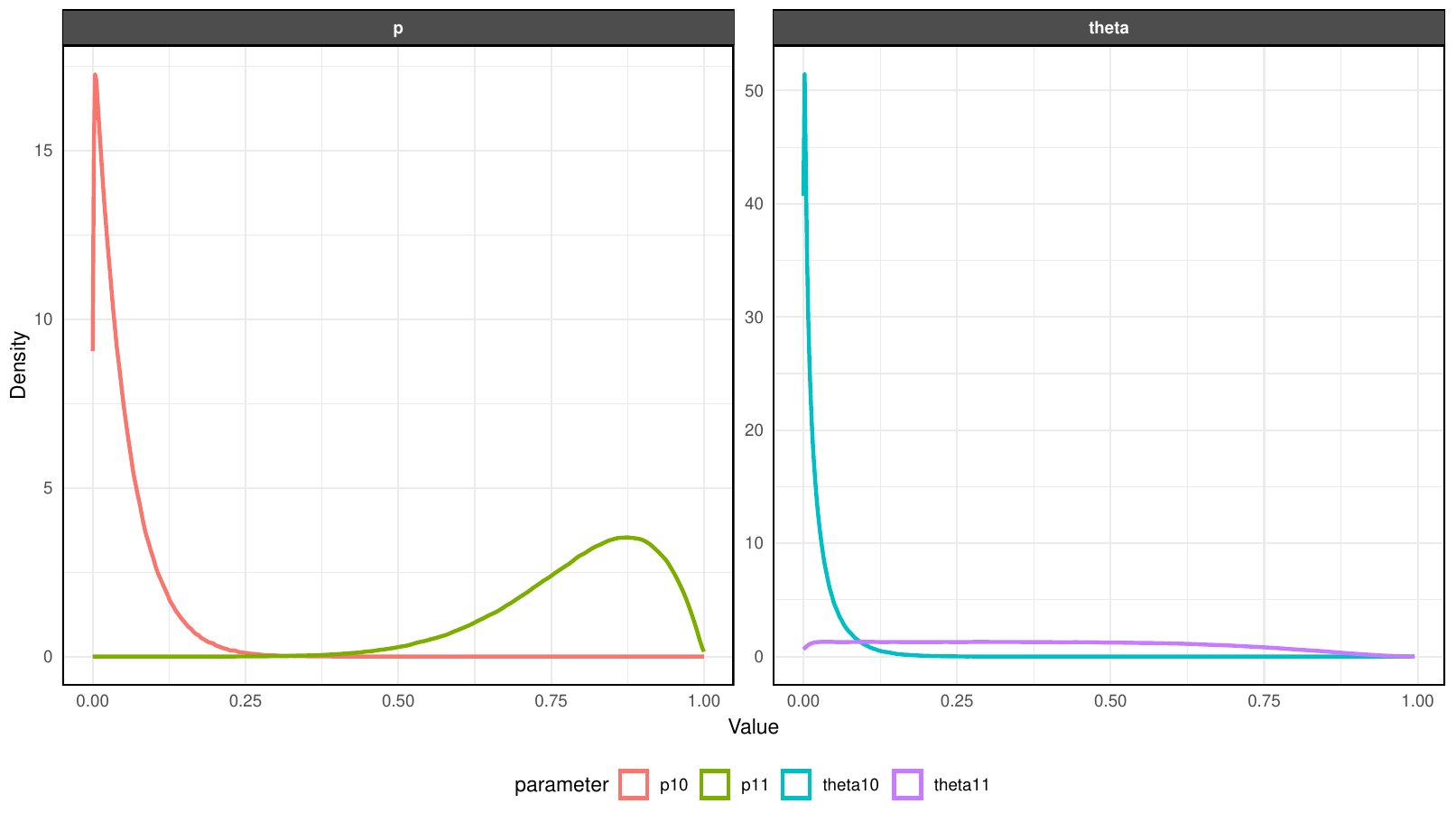}
        \caption{
        $p_{11}\sim \mathrm{Beta}(8,2)$,
        $p_{10}\sim \mathrm{Beta}(1,19)$, and $q \sim \mathrm{Beta}(1, 1)$.}
        \label{fig:sim-study-priors}
    \end{subfigure}
    \hspace{0.02\textwidth}
    \begin{subfigure}[t]{0.4\textwidth}
        \centering
        \includegraphics[width=\linewidth]
        {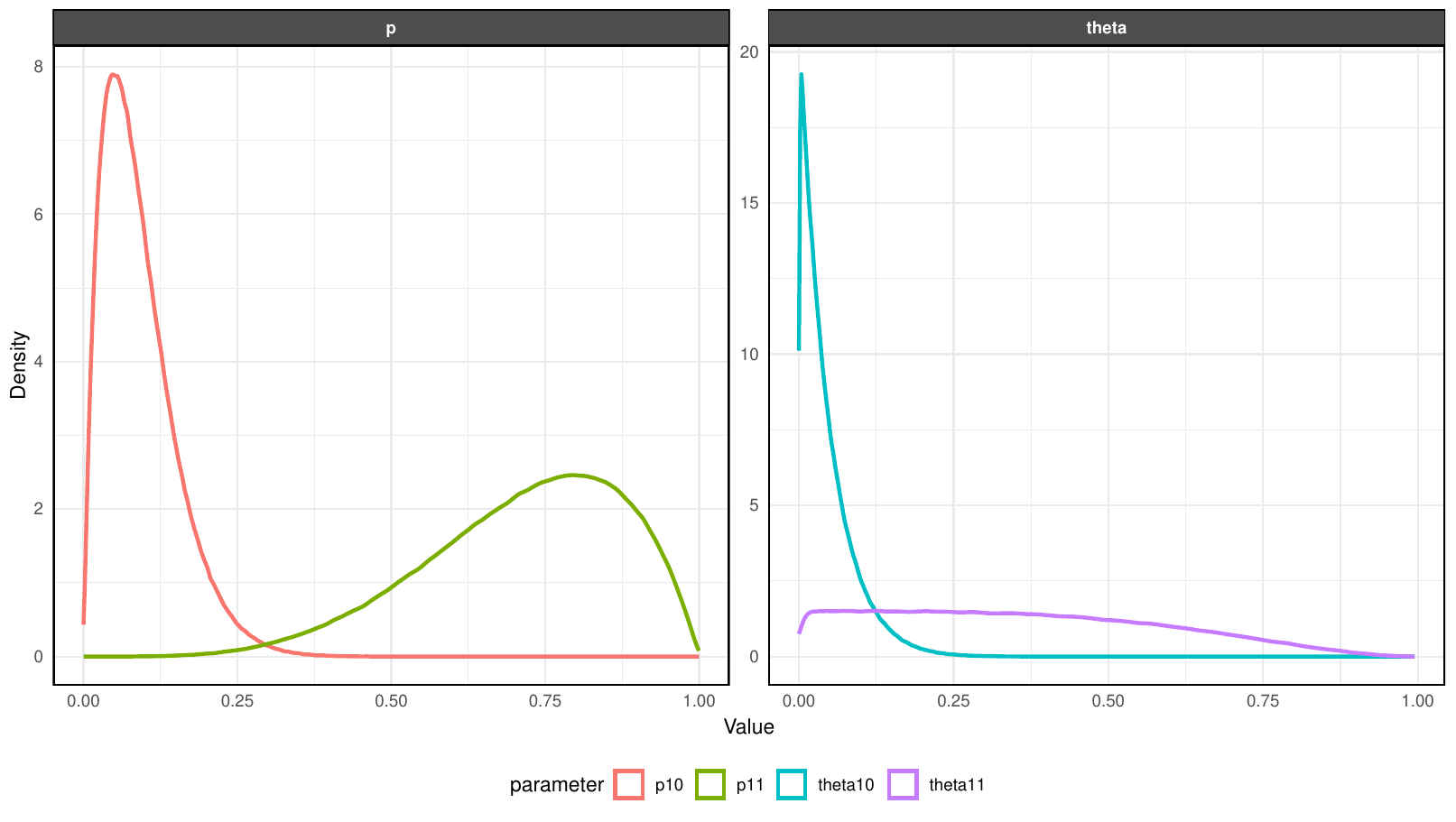}
        \caption{
        $p_{11}\sim \mathrm{Beta}(5,2)$,
        $p_{10}\sim \mathrm{Beta}(2,20)$, and $q \sim \mathrm{Beta}(1, 1)$.}
        \label{fig:case-study-priors}
    \end{subfigure}
    \caption{
    Induced prior distributions under different specifications for 
    $p_{11}$, $p_{10}$ and $q$. Panels (a), (b), and (c) show strongly
    informative priors that enforce clear separation between $p_{11}$
    and $p_{10}$ on the left and $\theta_{11}$ and
    $\theta_{10}$. Also, the non-informative prior on $q$ has a strong influence on $\theta_{11}$ since it becomes flat under 
    a weakly-informative or uniform prior for $q$. In Panel (c), with 
    weakly-informative priors for $p_{11}$, $p_{10}$, and an uniform 
    prior on $q$. There is strong overlap between $\theta_{11}$ and 
    $\theta_{10}$, which has repercussions for model identification. 
    Such weak priors result in $\hat{R}$ exceeding 1, for example, 
    1.05, indicating poor mixing. Panel (e) shows the priors for 
    $p_{11}$ and $p_{10}$ and the induced priors for $\theta_{11}$ and 
    $\theta_{10}$ used for the simulation study. Finally, Panel (f) 
    shows the priors for $p_{11}$ and $p_{10}$ and the induced priors 
    for $\theta_{11}$ and $\theta_{10}$ used for the case studies, i.e.,
    New Zealand and the United Kingdom.
    }
    
    \label{fig:induced-priors}
\end{figure}

\subsection{Simulation results}\label{supp:simstudy}
\begin{table}[H]
\centering
\caption{Performance metrics for the generalised linear model (GLM), 
the oracle model (Oracle), the two-visit detection/non-detection model 
(DN-2-M), the single-visit detection/non-detection model (DN-1-M), and 
the detection-only model (DO-M) with $q = 1$, under three prevalence 
scenarios: $\beta_0 = 0$ ($\bar{\psi} = 0.50$), $\beta_0 = 1.5$ 
($\bar{\psi} = 0.82$), and $\beta_0 = -1.5$ ($\bar{\psi} = 0.18$). 
Performance metrics are the mean relative bias (RB), mean root mean 
squared error (RMSE), and coverage (COV) of 95\% posterior credible 
intervals across 500 simulation replicates. RB is not reported for 
$\beta_0$ under Scenario 1 since the true value is zero.}
\label{tab:sim1}
\begin{tabular}{llrrrrrrrrr}
\hline
& & \multicolumn{3}{c}{$\beta_0 = 0$} 
& \multicolumn{3}{c}{$\beta_0 = 1.5$} 
& \multicolumn{3}{c}{$\beta_0 = -1.5$} \\
\cmidrule(lr){3-5}\cmidrule(lr){6-8}\cmidrule(lr){9-11}
Model & Parameter & RB & RMSE & COV & RB & RMSE & COV 
& RB & RMSE & COV \\
\hline
\multirow{4}{*}{GLM}
& $\beta_0$ & --- & 0.371 & 0.000 & -0.688 & 1.032 & 0.000 
& -0.083 & 0.136 & 0.374 \\
& $\beta_1$ & -0.399 & 0.480 & 0.000 & -0.578 & 0.695 & 0.000 
& -0.325 & 0.392 & 0.000 \\
& $\beta_2$ & -0.385 & 0.390 & 0.000 & -0.567 & 0.570 & 0.000 
& -0.329 & 0.335 & 0.004 \\
& $\beta_3$ & -0.402 & 0.403 & 0.000 & -0.572 & 0.573 & 0.000 
& -0.328 & 0.330 & 0.000 \\
\hline
\multirow{4}{*}{Oracle}
& $\beta_0$ & --- & 0.047 & 0.954 & 0.002 & 0.059 & 0.952 
& -0.001 & 0.059 & 0.958 \\
& $\beta_1$ & 0.002 & 0.041 & 0.968 & -0.002 & 0.050 & 0.948 
& 0.000 & 0.043 & 0.958 \\
& $\beta_2$ & -0.003 & 0.070 & 0.956 & -0.005 & 0.081 & 0.962 
& -0.007 & 0.073 & 0.970 \\
& $\beta_3$ & -0.001 & 0.039 & 0.956 & -0.000 & 0.046 & 0.954 
& 0.003 & 0.043 & 0.940 \\
\hline
\multirow{4}{*}{DN-2-M}
& $\beta_0$ & --- & 0.068 & 0.960 & 0.009 & 0.086 & 0.954 
& -0.002 & 0.080 & 0.952 \\
& $\beta_1$ & 0.002 & 0.049 & 0.958 & -0.001 & 0.066 & 0.962 
& 0.001 & 0.054 & 0.952 \\
& $\beta_2$ & -0.005 & 0.082 & 0.952 & -0.004 & 0.104 & 0.954 
& -0.008 & 0.085 & 0.960 \\
& $\beta_3$ & -0.002 & 0.047 & 0.960 & 0.003 & 0.061 & 0.952 
& 0.003 & 0.052 & 0.950 \\
\hline
\multirow{4}{*}{DN-1-M}
& $\beta_0$ & --- & 0.111 & 0.956 & 0.009 & 0.166 & 0.948 
& -0.011 & 0.142 & 0.946 \\
& $\beta_1$ & -0.002 & 0.109 & 0.964 & 0.008 & 0.119 & 0.962 
& -0.016 & 0.121 & 0.968 \\
& $\beta_2$ & -0.008 & 0.123 & 0.962 & 0.007 & 0.153 & 0.966 
& -0.031 & 0.133 & 0.952 \\
& $\beta_3$ & -0.002 & 0.095 & 0.970 & 0.012 & 0.104 & 0.962 
& -0.016 & 0.109 & 0.964 \\
\hline
\multirow{4}{*}{DO-M}
& $\beta_0$ & --- & 0.118 & 0.958 & 0.006 & 0.162 & 0.970 
& -0.013 & 0.147 & 0.942 \\
& $\beta_1$ & 0.001 & 0.104 & 0.976 & 0.005 & 0.115 & 0.984 
& 0.016 & 0.120 & 0.976 \\
& $\beta_2$ & -0.011 & 0.120 & 0.972 & -0.008 & 0.151 & 0.958 
& 0.011 & 0.133 & 0.972 \\
& $\beta_3$ & 0.002 & 0.094 & 0.960 & 0.006 & 0.100 & 0.974 
& 0.017 & 0.103 & 0.976 \\
\hline
\end{tabular}
\end{table}
\begin{table}[htbp]
\centering
\caption{Performance metrics for DO-M across six values of site visit 
probability $q \in \{0.5, 0.6, 0.7, 0.8, 0.9, 1.0\}$, under three 
prevalence scenarios: $\beta_0 = 0$ ($\bar{\psi} = 0.50$), 
$\beta_0 = 1.5$ ($\bar{\psi} = 0.82$), and $\beta_0 = -1.5$ 
($\bar{\psi} = 0.18$). Performance metrics are the mean relative bias 
(RB), mean root mean squared error (RMSE), and coverage (COV) of 95\% 
posterior credible intervals across 500 simulation replicates. RB is 
not reported for $\beta_0$ under Scenario 1 since the true value is 
zero.}
\label{tab:sim2}
\begin{tabular}{llrrrrrrrrr}
\hline
& & \multicolumn{3}{c}{$\beta_0 = 0$} 
& \multicolumn{3}{c}{$\beta_0 = 1.5$} 
& \multicolumn{3}{c}{$\beta_0 = -1.5$} \\
\cmidrule(lr){3-5}\cmidrule(lr){6-8}\cmidrule(lr){9-11}
$q$ & Parameter & RB & RMSE & COV & RB & RMSE & COV 
& RB & RMSE & COV \\
\hline
\multirow{4}{*}{0.5}
& $\beta_0$ & --- & 0.196 & 0.962 & -0.026 & 0.311 & 0.954 & -0.003 & 0.211 & 0.966\\
& $\beta_1$ & -0.008 & 0.159 & 0.984 & -0.012 & 0.183 & 0.978 & -0.032 & 0.175 & 0.978\\
& $\beta_2$ & -0.017 & 0.190 & 0.964 & -0.045 & 0.249 & 0.960 & -0.048 & 0.192 & 0.960\\
& $\beta_3$ & -0.007 & 0.142 & 0.978 & -0.012 & 0.160 & 0.980 & -0.029 & 0.155 & 0.990\\
\hline
\multirow{5}{*}{0.6} & $\beta_0$ & --- & 0.194 & 0.946 & -0.001 & 0.265 & 0.960 & -0.015 & 0.192 & 0.958\\
& $\beta_1$ & 0.019 & 0.162 & 0.974 & 0.007 & 0.164 & 0.986 & -0.025 & 0.152 & 0.976\\
& $\beta_2$ & -0.012 & 0.173 & 0.966 & -0.024 & 0.233 & 0.958 & -0.043 & 0.172 & 0.966\\
& $\beta_3$ & 0.008 & 0.135 & 0.982 & 0.005 & 0.149 & 0.978 & -0.036 & 0.139 & 0.970\\
\hline
\multirow{5}{*}{0.7} & $\beta_0$ & --- & 0.156 & 0.966 & -0.001 & 0.240 & 0.952 & -0.008 & 0.167 & 0.966\\
& $\beta_1$ & 0.006 & 0.144 & 0.976 & 0.003 & 0.153 & 0.976 & -0.019 & 0.150 & 0.978\\
& $\beta_2$ & -0.009 & 0.165 & 0.956 & -0.015 & 0.204 & 0.964 & -0.033 & 0.159 & 0.960\\
& $\beta_3$ & 0.011 & 0.131 & 0.964 & 0.002 & 0.127 & 0.978 & -0.018 & 0.134 & 0.974\\
\hline
\multirow{5}{*}{0.8} & $\beta_0$ & --- & 0.139 & 0.966 & -0.003 & 0.207 & 0.968 & -0.013 & 0.166 & 0.942\\
& $\beta_1$ & -0.008 & 0.136 & 0.974 & -0.001 & 0.139 & 0.980 & 0.003 & 0.138 & 0.968\\
& $\beta_2$ & -0.018 & 0.152 & 0.960 & -0.014 & 0.182 & 0.962 & -0.009 & 0.156 & 0.958\\
& $\beta_3$ & -0.009 & 0.121 & 0.958 & -0.003 & 0.119 & 0.984 & 0.001 & 0.123 & 0.970\\
\hline
\multirow{5}{*}{0.9} & $\beta_0$ & --- & 0.128 & 0.962 & 0.001 & 0.176 & 0.972 & -0.015 & 0.147 & 0.952\\
& $\beta_1$ & -0.002 & 0.117 & 0.972 & 0.002 & 0.127 & 0.986 & 0.010 & 0.125 & 0.986\\
& $\beta_2$ & -0.009 & 0.130 & 0.976 & -0.003 & 0.166 & 0.976 & -0.004 & 0.137 & 0.976\\
& $\beta_3$ & -0.000 & 0.104 & 0.968 & 0.005 & 0.105 & 0.984 & 0.008 & 0.114 & 0.978\\
\hline
\multirow{5}{*}{1} & $\beta_0$ & --- & 0.118 & 0.958 & 0.006 & 0.162 & 0.970 & -0.013 & 0.147 & 0.942\\
& $\beta_1$ & 0.001 & 0.104 & 0.976 & 0.005 & 0.115 & 0.984 & 0.016 & 0.120 & 0.976\\
& $\beta_2$ & -0.011 & 0.120 & 0.972 & -0.008 & 0.151 & 0.958 & 0.011 & 0.133 & 0.972\\
& $\beta_3$ & 0.002 & 0.094 & 0.960 & 0.006 & 0.100 & 0.974 & 0.017 & 0.103 & 0.976\\
\hline
\end{tabular}
\end{table}

\begin{table}[htbp]
\centering
\caption{Performance metrics for three cases of model misspecification, 
with DO-M results shown for $q = 1$. Case 1: an important covariate 
$x_3$ is omitted. Case 2: an incorrect covariate $x_3^*$ is used in 
place of $x_3$. Case 3: an irrelevant covariate $x_4$ is included. 
Performance metrics are the mean relative bias (RB), mean root mean 
squared error (RMSE), and coverage (COV) of 95\% posterior credible 
intervals across 500 simulation replicates.}
\label{tab:sim3}
\begin{tabular}{llrrrrrrrrr}
\hline
& & \multicolumn{3}{c}{Case 1} 
& \multicolumn{3}{c}{Case 2} 
& \multicolumn{3}{c}{Case 3} \\
\cmidrule(lr){3-5}\cmidrule(lr){6-8}\cmidrule(lr){9-11}
Model & Parameter & RB & RMSE & COV & RB & RMSE & COV 
& RB & RMSE & COV \\
\hline
\multirow{4}{*}{Oracle}
& $\beta_0$ & --- & 0.040 & 0.964 & --- & 0.040 & 0.966 
& --- & 0.047 & 0.954 \\
& $\beta_1$ & -0.157 & 0.192 & 0.000 & -0.157 & 0.191 & 0.000 
& 0.003 & 0.042 & 0.964 \\
& $\beta_2$ & -0.157 & 0.168 & 0.294 & -0.157 & 0.168 & 0.300 
& -0.002 & 0.070 & 0.956 \\
& $\beta_3$ & -1.003 & 1.003 & 0.000 & --- & --- & --- 
& -0.001 & 0.039 & 0.956 \\
& $\beta_4$ & --- & --- & --- & --- & --- & --- 
& 0.033 & --- & 0.962 \\
\hline
\multirow{4}{*}{DN-2-M}
& $\beta_0$ & --- & 0.071 & 0.954 & --- & 0.071 & 0.952 
& --- & 0.068 & 0.964 \\
& $\beta_1$ & -0.159 & 0.195 & 0.010 & -0.159 & 0.195 & 0.010 
& 0.002 & 0.049 & 0.960 \\
& $\beta_2$ & -0.160 & 0.174 & 0.430 & -0.159 & 0.174 & 0.436 
& -0.004 & 0.082 & 0.956 \\
& $\beta_3$ & -1.002 & 1.003 & 0.000 & --- & --- & --- 
& -0.001 & 0.047 & 0.958 \\
& $\beta_4$ & --- & --- & --- & --- & --- & --- 
& 0.039 & --- & 0.946 \\
\hline
\multirow{4}{*}{DN-1-M}
& $\beta_0$ & --- & 0.147 & 0.970 & --- & 0.149 & 0.968 
& --- & 0.112 & 0.960 \\
& $\beta_1$ & -0.167 & 0.221 & 0.784 & -0.152 & 0.207 & 0.842 
& 0.008 & 0.112 & 0.970 \\
& $\beta_2$ & -0.164 & 0.196 & 0.838 & -0.149 & 0.186 & 0.874 
& 0.002 & 0.125 & 0.956 \\
& $\beta_3$ & -1.002 & 1.003 & 0.000 & --- & --- & --- 
& 0.008 & 0.097 & 0.972 \\
& $\beta_4$ & --- & --- & --- & --- & --- & --- 
& 0.048 & --- & 0.954 \\
\hline
\multirow{4}{*}{DO-M}
& $\beta_0$ & --- & 0.142 & 0.968 & --- & 0.146 & 0.964 
& --- & 0.115 & 0.950 \\
& $\beta_1$ & -0.140 & 0.193 & 0.846 & -0.124 & 0.180 & 0.876 
& 0.018 & 0.112 & 0.976 \\
& $\beta_2$ & -0.136 & 0.174 & 0.878 & -0.120 & 0.165 & 0.904 
& 0.011 & 0.123 & 0.968 \\
& $\beta_3$ & -1.000 & 1.001 & 0.000 & --- & --- & --- 
& 0.016 & 0.103 & 0.962 \\
& $\beta_4$ & --- & --- & --- & --- & --- & --- 
& 0.050 & --- & 0.938 \\
\hline
\end{tabular}
\end{table}

\begin{table}[htbp]
\centering
\caption{Predictive performance of DO-M, the generalised linear model 
(GLM), Maximum entropy (Maxent), boosted regression trees (BRT), and 
random forest (RF) for $q = 0.5$ and $q = 1$. Metrics are the mean 
area under the receiver operating characteristic curve (AUC) and mean 
Brier score (BS) across 500 simulation replicates, computed against the 
true latent occupancy states $Z_i$. Posterior credible intervals for 
DO-M are shown in parentheses.}
\label{tab:sim4}
\begin{tabular}{llrrrrr}
\hline
$q$ & Metric & DO-M & GLM & Maxent & BRT & RF \\
\hline
\multirow{2}{*}{0.5}
& AUC & 0.8302 (0.828, 0.831) & 0.8305 & 0.8283 & 0.8231 & 0.8115 \\
& BS  & 0.3113 (0.305, 0.318) & 0.1955 & 0.1790 & 0.1975 & 0.1845 \\
\hline
\multirow{2}{*}{1.0}
& AUC & 0.8307 (0.830, 0.831) & 0.8309 & 0.8295 & 0.8271 & 0.8111 \\
& BS  & 0.1785 (0.177, 0.181) & 0.1786 & 0.1786 & 0.1799 & 0.1846 \\
\hline
\end{tabular}
\end{table}

\subsection{Additional simulation results}\label{supp:convergence}
\begin{figure}[H]
    \centering
    \begin{subfigure}{0.65\textwidth}
        \centering
        \includegraphics[width=\linewidth]{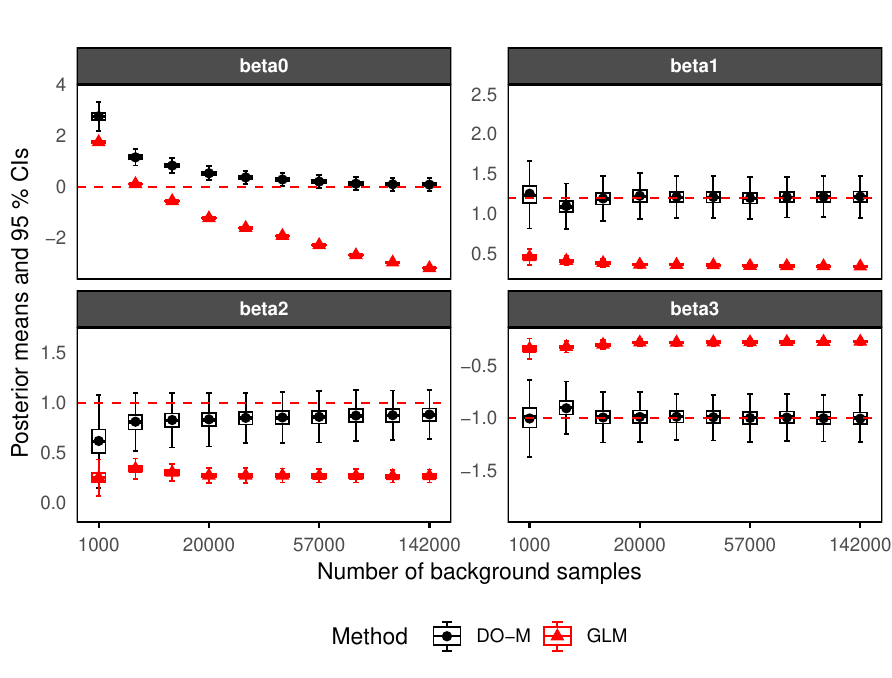}
   \caption{
   }
   \label{fig:supp_convergence_beta0}
    \end{subfigure}
    \vfill
    \begin{subfigure}{0.65\textwidth}
        \centering
        \includegraphics[width=\linewidth]{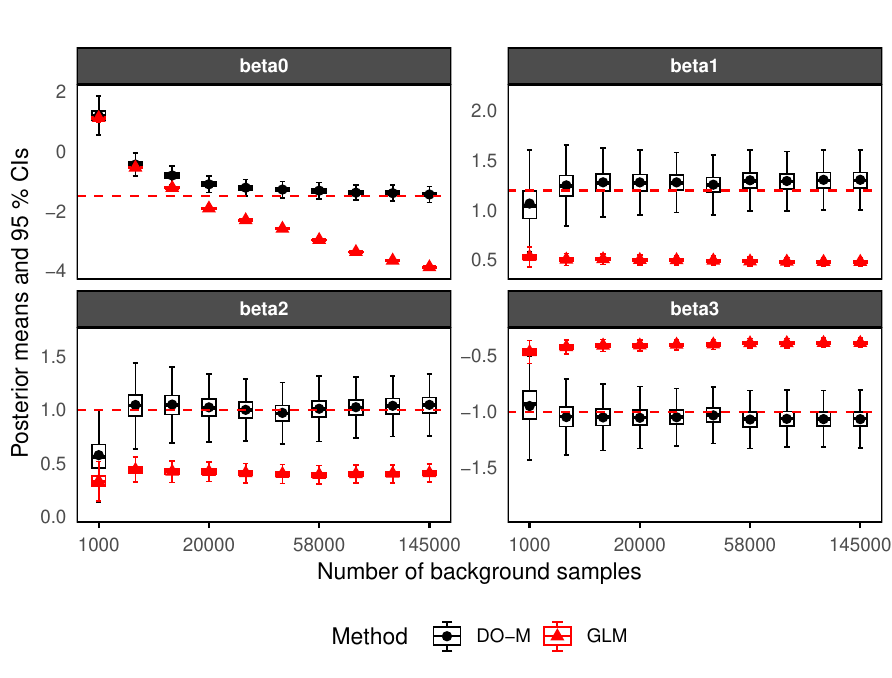}
   \caption{
   }
   \label{fig:supp_convergence_beta2}
    \end{subfigure}
    \caption{(a) The intercept and coefficient estimates for 
   DO-M and GLM as the background sample size grows, under Scenario 1 
   ($\beta_0 = 0$, $\bar{\psi} = 0.50$). (b) The intercept and coefficient estimates for 
   DO-M and GLM as the background sample size grows, under Scenario 3 
   ($\beta_0 = 1.5$, $\bar{\psi} = 0.82$).}
\end{figure}

\subsection{Descriptive statistics on Case study 1 (New Zealand)}\label{supp:casestudy1}
\begin{table}[H]
\centering
\caption{Case study 1: Description of environmental data.}
\label{tab:variables}
\centering
\begin{tabular}{lll}
  \hline
Variables &
  \multicolumn{1}{c}{Type} &
  \multicolumn{1}{c}{Description} \\ 
  \hline 
  \addlinespace
  Age & categorical & soil parent material: age since last
major rejuvenation \\
  Deficit & continuous &  mean October vapor pressure
deficit at 0900 hours   \\
  Dem & continuous & digital elevation model  \\
  Hillshade & continuous & surrogate for slope and aspect \\
  Mas & continuous & mean annual solar radiation  \\
  Mat & continuous & mean annual temperature \\
  R2pet & continuous & average monthly ratio of potential
evapotranspiration  \\
  Rain & continuous & average annual rainfall   \\
  Slope & continuous & slope   \\
  Sseas & continuous & solar radiation seasonality   \\
  Toxicats & categorical & toxic cations in soil    \\
  Tseas & continuous & temperature seasonality  \\
  Vpd & continuous & annual vapor pressure deficit   \\
  Latitude & continuous & latitude \\
   \hline
\end{tabular}
\end{table}

\begin{table}[H]
\centering
\caption{Case study 1: Summary statistics for continuous 
environmental variables. Categorical variables \textit{age} and 
\textit{toxicats} are excluded from this table. Continuous variables 
were standardised to have mean zero and unit variance prior to model 
fitting.}
\label{tab:descriptive}

\centering
\begin{tabular}{lllllll}
  \hline
Variables & \multicolumn{1}{c}{n} & \multicolumn{1}{c}{Mean} & \multicolumn{1}{c}{SD} & \multicolumn{1}{c}{Median} & \multicolumn{1}{c}{Min} & \multicolumn{1}{c}{Max} \\ 
  \hline
  \addlinespace
    Deficit & 19120 & 1.82 & 7.94 & 0.00 & 0.00 & 123.00 \\ 
    Dem & 19120 & 562.29 & 350.85 & 525.00 & 0.00 & 2020.00 \\ 
    Hillshade & 19120 & 169.56 & 48.96 & 178.00 & 0.00 & 254.00 \\ 
    Mas & 19120 & 1359.94 & 98.05 & 1373.00 & 1168.00 & 1547.00 \\ 
    Mat & 19120 & 94.25 & 20.49 & 93.00 & 19.00 & 158.00\\ 
    R2pet & 19120 & 83.75 & 51.78 & 60.00 & 20.00 & 282.00 \\ 
    Rain & 19120 & 3175.75 & 1716.48 & 2397.00 & 739.00 & 9157.00 \\ 
    Slope & 19120 & 19.46 & 9.95 & 20.00 & 19.45 & 54.00 \\ 
   Sseas & 19120 & 19.43 & 33.88 & 18.00 & -75.00 & 98.00 \\ 
   Tseas & 19120 & 43.14 & 102.89 & 33.00 & -236.00 & 336.00 \\ 
   Vpd & 19120 & 24.47 & 9.63 & 25.00 & 0.00 & 58.00 \\ 
   Latitude & 19120 & 46.85 &  1.98 & 46.35 & 43.00 & 51.73 \\
   \hline
\end{tabular}

\end{table}
\begin{table}[H]
\centering
\caption{Case study 1: Results for the single-visit 
detection/non-detection model (DN-1-M) applied to two anonymous 
vascular plant species from New Zealand. The mean, standard deviation 
(SD), and 95\% posterior credible intervals (CI) are reported for the 
intercept, coefficients of environmental conditions, and observation 
process parameters $p_{11}$ and $p_{10}$.}
\label{tab:nz_occ1}

\centering
\begin{tabular}[t]{lrrlrrl}
\toprule
\multicolumn{1}{c}{ } & \multicolumn{3}{c}{DN-1-M} & \multicolumn{3}{c}{DO-M} \\
\cmidrule(l{3pt}r{3pt}){2-4} \cmidrule(l{3pt}r{3pt}){5-7}
\multicolumn{1}{c}{Parameters} & \multicolumn{1}{c}{Mean} & \multicolumn{1}{c}{SD} & \multicolumn{1}{c}{95\% CI} & \multicolumn{1}{c}{Mean} & \multicolumn{1}{c}{SD} & \multicolumn{1}{c}{95\% CI}\\
\midrule
Intercept & 1.741 & 0.224 & (1.297, 2.175) & 1.744 & 0.226 & (1.300, 2.191)\\
Age0 & -0.110 & 0.714 & (-1.367, 1.482) & -0.108 & 0.725 & (-1.376, 1.491)\\
Age1 & 1.124 & 0.392 & (0.405, 1.954) & 1.128 & 0.394 & (0.392, 1.943)\\
Deficit & 0.943 & 0.316 & (0.298, 1.541) & 0.943 & 0.320 & (0.303, 1.552)\\
Dem & 2.643 & 0.137 & (2.378, 2.918) & 2.646 & 0.133 & (2.396, 2.914)\\
Hillshade & -0.030 & 0.037 & (-0.103, 0.043) & -0.030 & 0.036 & (-0.103, 0.041)\\
Mas & 2.350 & 0.342 & (1.685, 3.025) & 2.346 & 0.336 & (1.695, 3.009)\\
Rain & 1.502 & 0.115 & (1.280, 1.731) & 1.502 & 0.115 & (1.281, 1.730)\\
Slope & -1.411 & 0.077 & (-1.565, -1.263) & -1.412 & 0.077 & (-1.567, -1.264)\\
Sseas & 1.413 & 0.194 & (1.029, 1.795) & 1.411 & 0.190 & (1.040, 1.790)\\
Tseas & 0.015 & 0.089 & (-0.160, 0.191) & 0.013 & 0.089 & (-0.161, 0.188)\\
Latitude & -3.507 & 0.147 & (-3.806, -3.228) & -3.510 & 0.144 & (-3.810, -3.235)\\
$p_{11}$ & 0.512 & 0.007 & (0.498, 0.526) & --- & --- & --- \\
$p_{10}$ & 0.001 & 0.001 & (0.000, 0.003) & --- & --- & --- \\
$\theta_{11}$ & --- & --- & --- & 0.512 & 0.007 & (0.498, 0.525)\\
$\theta_{10}$ & --- & --- & --- & 0.001 & 0.001 & (0.000, 0.003)\\
\bottomrule
\end{tabular}

\end{table}

\begin{figure}[H]
    \centering
    \begin{subfigure}{0.8\textwidth}
        \centering
        \includegraphics[width=\linewidth]{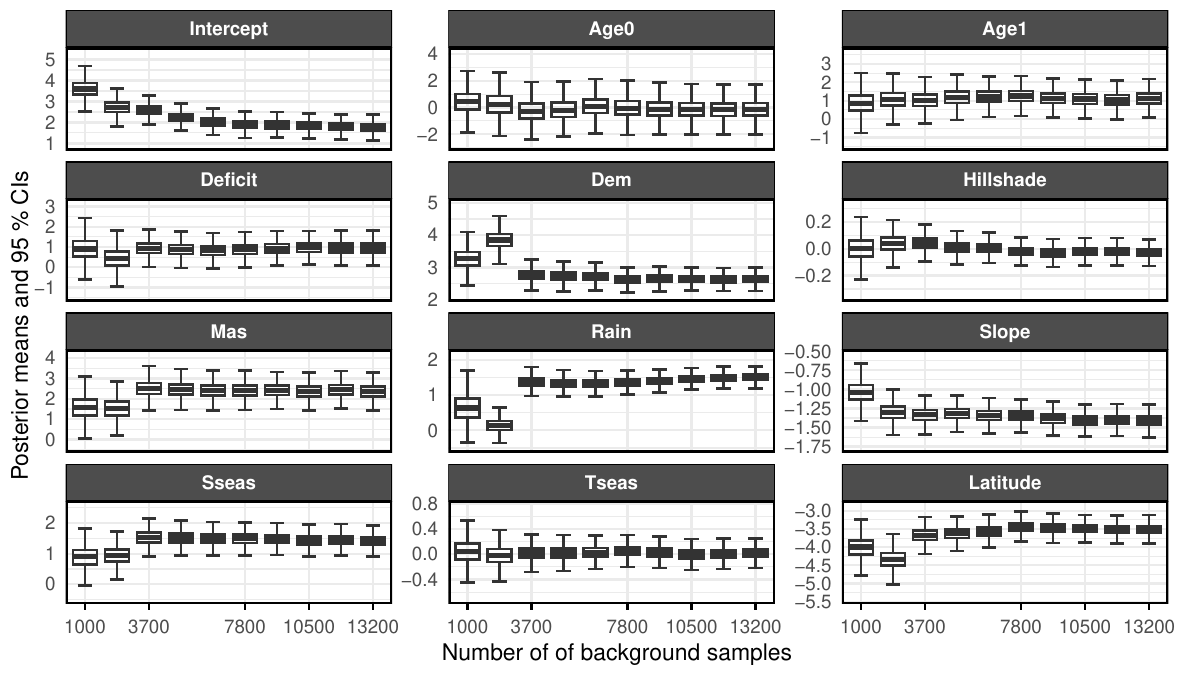}
   \caption{
   }
   \label{fig:supp_convergence_nz42}
    \end{subfigure}
    \vfill
    \begin{subfigure}{0.8\textwidth}
        \centering
        \includegraphics[width=\linewidth]{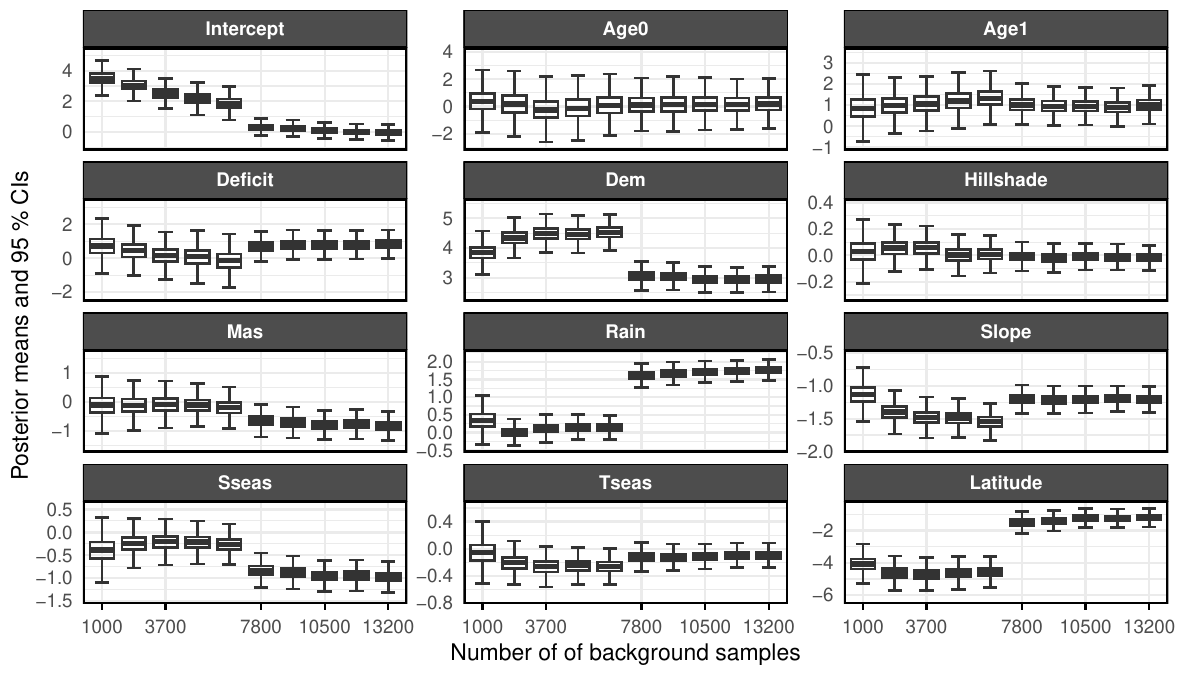}
   \caption{
   }
   \label{fig:supp_convergence_CS1}
    \end{subfigure}
    \caption{(a) Convergence of effect sizes and intercept estimates for 
    \textit{Case study 1} as background sample size grows when bio-climatic factors are latitude-adjusted, for example, Rain, Mas, Sseas, and Tseas. (b) Behaviour of effect sizes and intercept estimates under \textit{Case study 1} as background sample size grows, when the latitude variation is present in Rain, Mas, Sseas, and Tseas. Latitude is only added to the model as an explanatory variable.}
\end{figure}

\subsection{Descriptive statistics on Case study 2 (United Kingdom)}\label{supp:casestudy2}
\begin{table}[H]
\centering
\caption{Case study 2: Description of environmental data.}
\label{tab:variables2}

\centering
\begin{tabular}{lll}
\hline
Variables & \multicolumn{1}{c}{Type} & \multicolumn{1}{c}{Description} \\ 
\hline
\addlinespace
Rainfall & Continuous & Mean rainfall during growing season \\
Temperature & Continuous & Mean temperature during growing season \\
Dry Nitrogen & Continuous & Mean nitrogen deposition under dry conditions \\
Wet Nitrogen & Continuous & Mean nitrogen deposition from precipitation \\
Genome size & continuous & Total amount of DNA contained in a single complete genome \\
Bog & Categorical &  Bog \\
Heather & Categorical &  Heath grasslands \\
Acid/Neutral Grassland & Categorical &  Acid and Neutral Grasslands (Pasture) \\
Improved Grassland & Categorical & Improved Grassland (Meadow)  \\
Rocky & Categorical &  Rocky \\
Coniferous Woodland & Categorical & Coniferous Woodland  \\
Arable Horticulture & Categorical & Arable and horticulture  \\
Urban/Suburban & Categorical &  Urban and suburban land use types \\
Sediment Coastal & Categorical & Sediment Coastal  \\
Broad Leaved Woodland & Categorical & Broad Leaved Woodland (Forest)  \\
Calcarious Grassland & Categorical &  Calcarious Grassland \\

\hline
\end{tabular}
\end{table}
\begin{table}[H]
\centering
\caption{Case study 2: Summary statistics for continuous 
environmental variables. Continuous variables that were standardised to 
have mean zero and unit variance prior to model fitting.}
\label{tab:descriptive2}

\centering
\begin{tabular}[t]{lrrrrrr}
\toprule
Variable & n & Mean & Median & SD & Min & Max\\
\midrule
Rainfall & 2794 & 74.636 & 66.061 & 27.905 & 38.998 & 201.476\\
Temperature & 2794 & 11.898 & 11.993 & 1.538 & 6.473 & 15.187\\
Dry Nitrogen & 2794 & 3.470 & 3.464 & 2.022 & 0.248 & 12.257\\
Wet Nitrogen & 2794 & 7.243 & 7.345 & 3.549 & 0.903 & 25.644\\
Genome size & 2794 & 2.597 & 2.624 & 0.205 & 2.171 & 2.988\\
\bottomrule
\end{tabular}

\end{table}

\bibliographystyle{apalike}
\bibliography{references.bib}


\end{document}